\RequirePackage{silence} 
\documentclass[sn-mathphys]{sn-jnl}% Math and Physical Sciences Reference Style
\usepackage{color}

\usepackage{bbm}
\usepackage{amsmath}
\usepackage{amsfonts}
\usepackage{amssymb}
\DeclareMathAlphabet{\mathpzc}{OT1}{pzc}{m}{it}
\newcommand{\RR}{\mathbb{R}}
\newcommand{\ZZ}{\mathbb{Z}}
\newcommand{\PP}{\mathbb{P}}
\newcommand{\EE}{\mathbb{E}}
\newcommand{\TT}{\mathbb{T}}

\newcommand{\J}{ \Sigma}
\newcommand{\HH}{ \Omega}

\renewcommand{\x}{\boldsymbol{x}}{}
{}
{}

\renewcommand{\v}{{\hat y}}
\renewcommand{\u}{{\hat x}}

\renewcommand{\z}{{e}}

\renewcommand{\tw}{\tilde}{}
\renewcommand{\p}{\partial}{}

\newcommand{\dd}{\mathrm{d}}{}
\newcommand{\id}{\mathbbm{1}}{}
\newcommand{\ep}{\varepsilon}{}

\newcommand{\rr}{\Rho_r}
\newcommand{\bb}{\Rho_b}

\newcommand{\rrr}{P_\s}

\newcommand{\s}{\sigma}
\newcommand{\os}{{\bar \sigma}}

\newcommand{\beq}{\begin{equation}}
\newcommand{\eeq}{\end{equation}}
\newcommand{\lap}{\Delta^1}

\newcommand{\Rho}{{\cal P}}

\newcommand{\bfA}{{\bf A}}
\newcommand{\bfB}{{\bf B}}
\newcommand{\bfC}{{\bf C}}
\newcommand{\ind}{\id_{\{ \vert \v\vert =1  \}}}

\newcommand{\bfkappa}{\boldsymbol{\kappa}}

\newcommand{\bgfpsi}{\boldsymbol{\psi}}
\newcommand{\bgfpsitw}{\tilde{\boldsymbol{\psi}}}
\newcommand{\gfpsi}{{\psi}}
\newcommand{\gfpsitw}{\tilde{{\psi}}}

\newcommand{\zdstar}{{\mathbb{Z}_*^d}}
\newcommand{\zdstart}{{\mathbb{Z}_*^{2d}}}
\newcommand{\xbi}{\xi}

\newcommand{\xmid}{{\hat x}}
\newcommand{\ymid}{{\hat y}}
\newcommand{\zmid}{{\hat z}}

\newcommand{\xin}{{\bar x}}
\newcommand{\yin}{{\bar y}}
\newcommand{\zin}{{\bar z}}

\newcommand{\xx}{y}
\newcommand{\xxx}{z}

\newcommand{\ta}{g}

\PassOptionsToPackage{unicode}{hyperref}
\PassOptionsToPackage{naturalnames}{hyperref}
\jyear{2021}%

\theoremstyle{thmstyleone}%
\theoremstyle{thmstyletwo}%

\theoremstyle{thmstylethree}%

\begin{document}

\title[Two-species exclusion process via matched asymptotics]{Macroscopic behaviour in a two-species exclusion process via the method of matched asymptotics}

% GivenName	-> \fnm{Joergen W.}

\author*{\fnm{James} \sur{Mason}$^1$}\email{jm2386@cam.ac.uk}

\author{\fnm{Robert L} \sur{Jack}$^{1,2}$}%\email{rlj22@cam.ac.uk}

\author{\fnm{Maria} \sur{Bruna}$^1$}%\email{bruna@maths.cam.ac.uk}

\affil{$^1$ \orgdiv{Department of Applied Mathematics and Theoretical Physics}, \orgname{University of Cambridge}, \orgaddress{\street{Wilberforce Road}, \city{Cambridge}, \postcode{CB3 0WA}, \country{UK}}}

\affil{$^2$ \orgdiv{Yusuf Hamied Department of Chemistry}, \orgname{University of Cambridge}, \orgaddress{\street{Lensfield Road}, \city{Cambridge}, \postcode{CB2 1EW}, \country{UK}}}


\abstract{
We consider a two-species simple exclusion process on a periodic lattice. We use the method of matched asymptotics to derive evolution equations for the two population densities in the dilute regime, namely a cross-diffusion system of partial differential equations for the two species' densities. 
First, our result captures non-trivial interaction terms neglected in the mean-field approach, including a non-diagonal mobility matrix with explicit density dependence. 
Second, it generalises the rigorous hydrodynamic limit of Quastel [Commun. Pure Appl. Math. 45(6), 623--679 (1992)], valid for species with equal jump rates and given in terms of a non-explicit self-diffusion coefficient, to the case of unequal rates in the dilute regime. 
In the equal-rates case, by combining matched asymptotic approximations in the low- and high-density limits, we obtain a cubic polynomial approximation of the self-diffusion coefficient that is numerically accurate for all densities. This cubic approximation agrees extremely well with numerical simulations. It also coincides with the Taylor expansion up to the second-order in the density of the self-diffusion coefficient obtained using a rigorous recursive method.
}




\maketitle

\section{Introduction}
\label{sec:intro}

Stochastic models describing systems of interacting particles are widely used across many disciplines \cite{stochbio,reviewcanerbio,traffic}. A particular class of models concerns excluded-volume or steric interactions, which are local and remain strong regardless of the number of particles in the system. This is in stark contrast with weak or mean-field interactions, which are long-range and weaker as the number of particles increases. 
Excluded-volume interactions arise from constraints that forbid overlap of particles and, as a result, are particularly relevant in biological applications with crowding, such as transport of tumour cells \cite{tumorcellmigration}, wound healing \cite{woundhealing}, and predator-prey systems \cite{predatorprey}.  
Models for diffusive systems with excluded-volume interactions can be broadly split into continuous and discrete depending on the stochastic process employed to describe the motion of particles. The continuous approach considers correlated Brownian motions, with correlations appearing either through soft short-ranged interaction potentials or hard-core potentials representing particle shapes. The discrete approach consists of random walks on discrete spaces (e.g. a regular lattice) with exclusion rules that constrain the number of particles allowed in each lattice site. This paper focuses on one such model, namely a simple exclusion process (SEP) whereby only one particle is allowed per site. In particular, we study a SEP with two species of particles on a $d$-dimensional lattice, for $d \geq 2$.\footnote{In the case $d=1$ the system is a reducible Markov chain, as particles cannot jump past one another. This leads to drastically different sub-diffusive behaviour \cite{Liggbook}.}  The lattice spacing is $h$ and the total number of particles is $N$.

We focus on systems with many particles, in a limit where \textcolor{black}{the number of particles $N\to \infty$ and the lattice spacing $h\to0$.} 
The central object of interest is the macroscopic density, whose evolution can be described by partial differential equations (PDEs).
For a single-species SEP, the continuum model for the particle density $\rho(x,t)$ takes the form \cite{singleasep}
\begin{equation} \label{limit_single}
	\partial_t \rho = \nabla \cdot \left[ D \nabla \rho + \rho(1-\rho) \nabla V \right],
\end{equation}
where the diffusivity $D$ and the potential force $V$ are determined from the underlying asymmetric hopping rates of the process. The first approach to obtain the drift-diffusion equation \eqref{limit_single} is through a hydrodynamic limit, where the (random) empirical density converges in probability to the (deterministic) solution of the PDE \cite{Varbook,Liggbook, Bertini}, taking \textcolor{black}{the number of particles} $N\to \infty$ and \textcolor{black}{the lattice spacing }$h\to0$ together, keeping the occupied volume fraction $\phi = \int \rho \mathrm{d} x$ fixed.
An alternative method for deriving macroscopic evolution PDEs such as  \eqref{limit_single} is to analyse the average density directly \cite{stevens1997aggregation}, thus sidestepping the technical challenges associated with the convergence of random variables. The starting point of this approach is to consider the master equation for the joint density of the $N$ particle system \cite{baker2010microscopic}, from which a hierarchy of equations for the moments can be obtained (analogously to the BBGKY hierarchy \cite{BBGKY}). The macroscopic behaviour is determined by the first moment (mean density) equation, which in general (for interacting processes) depends on the second moment. A standard route to obtain a closed equation is to use a moment closure approximation such as a mean-field closure \cite{mean1,mean2}. 
Finally, the corresponding PDE description is found by performing a simple perturbation approach (taking the continuous limit assuming the density to be slowly varying \cite{kGEP}). 

The case of a single species of SEP (including asymmetric rates) is well understood:  the macroscopic limit obtained formally via a mean-field closure, coincides with the rigorous hydrodynamic limit \eqref{limit_single} studied in \cite{singleasep}. 
In \cite{Quastel} \textcolor{black}{and \cite{Quastel1999}}
, Quastel derived the hydrodynamic limit \textcolor{black}{and large deviations respectively, for finitely many} \emph{coloured species} undergoing a symmetric SEP (SSEP), corresponding to equal diffusivity $D$ amongst species and setting $V\equiv0$ in \eqref{limit_single}. The hydrodynamic equations in this case take the form of a cross-diffusion system of PDEs for the two species' densities $\rho_r, \rho_b$, 
\begin{equation}
	\p_t \begin{pmatrix}
	\rho_r \\
	\rho_b
	 \end{pmatrix} 
	 = \nabla \cdot \left[  M(\rho_r,\rho_b) \nabla \begin{pmatrix}
	\delta_{\rho_r} E\\
	\delta_{\rho_b}E 
\end{pmatrix} \right],
\label{equ:gen-hydro}
\end{equation}
where $M$ is a mobility matrix and $\delta_{\rho}E$ denotes the functional derivative of a suitable free-energy functional $E$. The form \eqref{equ:gen-hydro} can be indentified as a 2-Wasserstein gradient flow \cite{Burger, Peletier}. The mobility matrix $M$ in \cite{Quastel} is full and depends on the self-diffusion coefficent $D_s(\rho)$ of a tagged particle in a SSEP at uniform density $\rho = \rho_r + \rho_b$.
This $D_s(\rho) \in C^\infty([0,1])$ \cite{lamdim}, but its dependence on $\rho$ is not known, although it can be determined from a variational formula \cite{Sophn}.  In contrast to the single-species SEP, analysis of this hydrodynamic limit is challenging for multi-species systems, because they are not of gradient type \cite{Varbook}, this technical distinction will be discussed below. 
Quastel's approach has been generalised to study an active SEP (where the orientation of the particle governs the asymmetry in the hopping rates) in \cite{Erignoux}. \textcolor{black}{The hydrodynamic limit has also been rigorously derived for similar multi-species systems in \cite{gabrielli1999onsager} and \cite{seo2018scaling}, for the zero-range process and interacting Brownian motions in one dimension, respectively. In both of these models the limiting macroscopic PDE is known explicitly.} 

For the two-species SEP, the master equation and mean-field closure approach~\cite{Burger,mean1} yields a result of the form \eqref{equ:gen-hydro} with an explicit diagonal mobility matrix.  Clearly, this result is not consistent with the rigorous hydrodynamic limit whose mobility matrix is full: this situation may be contrasted with the single-species case where the result of the mean-field closure is exact.

In this work, we derive a cross-diffusion system of the form \eqref{equ:gen-hydro} from the master equation using the method of matched asymptotics \cite{holmes2012introduction} in the low-density limit, $\phi\ll 1$. In contrast to the mean-field closure, our approach has the advantages of being systematic and consistent with the rigorous hydrodynamic limit \cite{Quastel}. In particular, 
we show that our mobility matrix $M$ agrees up to the expected order in $\phi$ with the result by Quastel \cite{Quastel}. Moreover, our approach provides an explicit mobility matrix, which makes the analysis of the PDE more amenable, and extends the result of Quastel \cite{Quastel} to handle the general case of different rates amongst species. This is the first result for two species with unequal rates that does not rely on a mean-field approximation to the authors' knowledge.
 
The method of matched asymptotics also yields predictions for the motion of a single (tracer) particle in these mixtures via its self-diffusion constant. For the case where the species have equal diffusivity, this yields the same result as the (rigorous) recursive computation of the self-diffusion constant of Landim \emph{et al.} \cite{lamdim}. In this case, we also consider an expansion around the high-density limit, which we combine with the low-density expansion to obtain a cubic approximation of the self-diffusion coefficient that performs well for the whole range of densities.  This leads to an explicit PDE for the density, which agrees well with numerical data.

The remainder of the paper is organised as follows. In Sec.~\ref{sec:mainresults} we define the model, discuss existing results for the coloured case, and summarise our main results. The derivation of a cross-diffusion system of the form \eqref{equ:gen-hydro} for the general case via mean-field and the method of matched asymptotic is given in Sec.~\ref{sec:matched}.  Then, Sec.~\ref{sec:self-diff} concerns the analysis of the self-diffusion coefficient $D_s$, and Sec.~\ref{sec:numerics} compares numerical simulations of the PDE systems with stochastic simulations of the microscopic model.  We draw together our conclusions in Sec.~\ref{sec: discussion}.

%%%%%%%%%%%%%%%%%%%%%%%%%%%%%%%%%%%%%%
\section{Model definitions and summary of main results}
\label{sec:mainresults}
%%%%%%%%%%%%%%%%%%%%%%%%%%%%%%%%%%%%%%

%%%%%%%%%%%%%%%%%%%%%%%%%%%%%%%%%%%%%%
\subsection{Microscopic model}
\label{sec:model}
%%%%%%%%%%%%%%%%%%%%%%%%%%%%%%%%%%%%%%

 Consider $N$ particles on a $d$-dimensional (hyper)-cubic lattice $\Omega$ with $L^d$ sites and $d \geq 2$.  We embed $\Omega$ in the unit $d$-dimensional torus $\TT^d$ by setting the lattice spacing $h=(1/L)$.  We impose a simple exclusion constraint so each site contains at most one particle.  Hence the lattice spacing, $h$, can be thought of as the particle diameter, and the lattice sites are points $x\in \Omega\subset \TT^d$ such that $(x/h)\in \{1,2,\dots,L\}^d$.  

We consider a system with two types (species) of particles, where the hopping rates depend on the species. We denote the species as `red' and `blue'. The total number of particles is $N$, of which $N_r$ are red, and $N_b$ are blue. Throughout, we use $\s \in \{ r, b\}$ to label a species and $\os$ for the opposite species.
 
The (random) configuration of species $\s$ at time $t$ is denoted by $\eta^{\s}_t \in \{ 1, 0 \}^\Omega$, where $\eta^{\s}(x) = 1$ if the site $x \in \HH$ is occupied by a $\s$-particle and $\eta^{\s}(x) = 0$ otherwise. The configuration of the whole system at time $t$ is denoted by $\eta_t = ( \eta^{r}_t, \eta^{b}_t  )\in \Sigma$, where $\Sigma = \{ (0,0), (1,0), (0,1) \}^\HH$.
 
The system evolves as a simple exclusion process (SEP), which is a Markov jump process on ${\J}$: a particle of species $\s$ at site $x$  attempts to jump to an adjacent site $y$ with rate $\lambda_\s (x,y)$; if the destination site is empty then the jump is executed; otherwise, the particle remains at $x$. 
The microscopic hopping rates $\lambda_\s$ are given by
\begin{equation}
	\lambda_\s (x,y) = \frac{D_\s}{h^2}\exp \left( \frac{V_\s(x) - V_\s(y)}{2} \right) \id_{ \{ \vert x-y\vert  =h \} }  , 
	\label{equ:rates-lambda-i}
\end{equation}
where $D_\s$ is a diffusion constant and $V_\s$ is a smooth potential, and $\vert \cdot \vert$ denotes the standard Euclidean norm. Since these rates respect detailed balance, the process is reversible with  stationary measure
\begin{equation}
	\pi ( \eta ) \propto   \exp \Big( - \sum_{x,\s} V_\s(x) \eta^\s(x) \Big),
\end{equation}
where the constant of proportionality is fixed by normalisation. 

The choice of lattice spacing $h$ and rate of order $h^{-2}$ corresponds to parabolic scaling; this ensures that, when taking $h\to0$, a single $\s$-particle's motion converges to a drift-diffusion process with drift $-D_\s \nabla V_\s$.\footnote{We take the convention that a diffusion process, $X$, with diffusion coefficient $D$ has the SDE $\text{d}X_t = \sqrt{2D}\text{d}B_t$ where $B$ is a standard Brownian motion. The probability density for $X_t$ evolves according to the PDE $\p_t p = D \nabla^2 p$.} (Note that particle jumps have always $\vert x-y\vert =h$ so $ V_i(x) - V_i(y)$ will be of order $h$.)
For a system with configuration $\eta = (\eta^r, \eta^b)$, a particle at site $x$ jumps to site $y$ with rate
\begin{equation}
	c(\eta, x,y) = \left[ 1-\eta^r(y) -\eta^b(y)\right] \left[
			\eta^r(x) \lambda_r(x,y) +\eta^b(x) \lambda_b(x,y) 
		\right].
	\label{equ:lambda-lambda}
\end{equation}
The \emph{discrete} density of particles of species $\s$ is described by
\begin{equation}
	\Rho_\s (x,t) = \EE\big[\eta^\s_t (x)\big].
	\label{equ:latt-rho}
\end{equation}
The value of $\Rho_\s (x)$ gives the probability that a lattice site $x$ is occupied by a $\s$ particle, so that $\sum_x \Rho_\s(x) = N_\s$. It is also convenient to define the total density
\begin{equation}
\Rho= \rr + \bb   .
\end{equation}
In the hydrodynamic limit, $N\to\infty$ and $h\to0$ at fixed volume fractions $\phi_\s=N_\s h^{d}$ we assume in the following that
\begin{alignat}{3}
	\Rho_\s \to \rho_\s, & \quad \quad & \Rho \to \rho = \rho_r + \rho_b,
\end{alignat}
where $\rho_\s$ and $\rho$ are the continuous densities in $\TT^d$ that appear in equations such as~\eqref{equ:gen-hydro}.

\subsection{Existing results for \texorpdfstring{$D_r=D_b$}{DreqDb}}
\label{sec:existing}

Hydrodynamic limits of simple exclusion processes for mixtures were first analysed by Quastel in \cite{Quastel}, with a recent extension by Erignoux in \cite{Erignoux}. Both these works consider the case where the diffusivity $D_\s$ is independent of $\s$. We briefly review their main results.
 
Consider the process described in Section \ref{sec:model} with  $D_r=D_b$ and $V_r=0=V_b$, that is, two `coloured' species undergoing a \emph{symmetric} simple exclusion process (SSEP). Without loss of generality, we choose $D_r=D_b=1$. The hydrodynamic limit of this process was analysed by \cite{Quastel}, who proved convergence of the (random) empirical densities $\varrho_t^\s = h^d \sum_{x \in \Omega} \eta_t^\s(x) \delta_x$
to deterministic densities $\rho_\s(x,t)$ solving a cross-diffusion system of  PDEs of the form \eqref{equ:gen-hydro}. In particular, the free energy is
\begin{subequations}
	\label{quastel}
	\begin{equation}
	\label{equ:E0}
		E_0[\rho_r,\rho_b] = \int_{\TT^d} \left[ \rho_r \log \rho_r + \rho_b \log \rho_b + (1 - \rho)\log(1-\rho)  \right] \dd x,
	\end{equation}
and the mobility matrix is
	\begin{equation}
	M^\text{sym}(\rho_r,\rho_b) = \frac{1-\rho}{\rho} \begin{pmatrix}
	{\rho_r^2}{}
	&&
	{\rho_r \rho_b}{}
	\\
	 {\rho_r\rho_b}{}
	 &&
	 {\rho_b^2}{}				
	 \end{pmatrix}
	 +
	 \frac{\rho_r\rho_b}{\rho}
	 D_s(\rho) 
	 \begin{pmatrix} 1 & -1 \\ -1 & 1 
	 \end{pmatrix}.
	 \label{equ:M-quas}
\end{equation}
\end{subequations}
where $D_s(\rho)$ is the self-diffusion constant of a single tagged particle in a symmetric simple exclusion process at density $\rho$.
(The dependence of $D_s$ on $\rho$ is not known in general, but it does have a variational characterisation \cite{lamdim}, we return to this object in later sections.) 

As discussed in \cite{Erignoux} the method can be extended to include a weak species-dependent drift. The expected generalisation to $V_r,V_b\neq0$ is that $M$ remains the same, while $E$ is replaced by \cite{Burger}
\begin{equation}
\label{equ:Etot}
E[\rho_r,\rho_b]  = E_0[\rho_r,\rho_b]  +  \int_{\TT^d} \left[ \rho_r V_r + \rho_b V_b   \right] \dd x  .
\end{equation}
The corresponding thermodynamic force is:
\begin{equation}
	\label{equ:Etot_expanded}
	\nabla \frac{\delta E}{\delta \rho_\s} = \frac{\nabla \rho_\s}{\rho_\s}+ \frac{\nabla \rho}{1- \rho} + \nabla V_\s  .
\end{equation}
[This thermodynamic force is the object $\nabla(\delta_{\rho_\sigma} E)$ that appears in \eqref{equ:gen-hydro}.]

In contrast to the hydrodynamic limit of a single species SEP, the limit for mixtures \cite{Quastel, Erignoux} is much more challenging. This is because they are \emph{nongradient systems}, meaning that the instantaneous particle currents along an edge cannot be written as a discrete gradient. The core tool to obtain hydrodynamic limits in gradient systems, namely an integration by parts, is unavailable for nongradient systems. Thus terms of \textcolor{black}{$O(h^{-1})$} need to be controlled by other means \cite{Quastel,Varbook}. 
A key advantage of the method of matched asymptotics used here is that it can deal with both gradient and nongradient systems in a relatively straightforward way, including the case of a mixture with unequal and space-dependent rates, which is not covered by the existing hydrodynamic limits literature. 

%%%%%%%%%%%%%%%%%%%%%%%%%%%%%%%
\subsection{Summary of results of this work}
\label{sec:summary}
%%%%%%%%%%%%%%%%%%%%%%%%%%%%%%%%

This work uses the method of matched asymptotics to analyse the time evolution of the mean density, including the general case with $D_r\neq D_b$ and $V_r, V_b\neq0$. We apply the method of matched asymptotics in the limit of low but finite volume fraction, corresponding to $\rho\ll1$ (but with arbitrary values of $\rho_r/\rho$ and $\rho_b/\rho$).  
We emphasise that the method yields an equation for the average density, but it does not establish that the (random) empirical density converges to this average value in the hydrodynamic limit.

\paragraph{Cross-diffusion system for \texorpdfstring{$D_r\ne D_b$}{DrneqDb} in the dilute regime}
We derive the following cross-diffusion system for the average densities $\rho_r,\rho_b$ through a systematic asymptotic expansion as $\rho = \rho_r + \rho_b \ll 1$, valid up to $O(\rho^2)$
\begin{subequations} \label{full_mainresult}
	\begin{equation}
	\p_t \begin{pmatrix}
	\rho_r \\
	\rho_b
	 \end{pmatrix} 
	 = \nabla \cdot \left[  M^\text{low}(\rho_r,\rho_b) \nabla \begin{pmatrix}
	\delta_{\rho_r} E\\
	\delta_{\rho_b}E 
\end{pmatrix} \right],
\label{gradflow}
\end{equation}
with free energy
	\begin{equation}
	\label{equ:Efm}
		E = \int_{\TT^d} \left[ \rho_r \log \rho_r + \rho_b \log \rho_b + (1 - \rho)\log(1-\rho) +  \rho_r V_r + \rho_b V_b \right] \dd x,
	\end{equation}
and mobility
\begin{equation}  
M^\text{low} =\begin{pmatrix} D_r & 0\\ 0& D_b \end{pmatrix} \left[ \frac{1-\rho}{\rho} \begin{pmatrix}
	{\rho_r^2}{}
	&&
	{\rho_r \rho_b}{}
	\\
	 {\rho_r\rho_b}{}
	 &&
	 {\rho_b^2}{}				
	 \end{pmatrix}
	 +
	 \frac{\rho_r\rho_b}{\rho}
	 \begin{pmatrix} \mu_r(\rho) & -\mu_b(\rho)  \\ -\mu_r(\rho)  & \mu_b(\rho)  
	 \end{pmatrix} \right],
	 \label{equ:M-mix}
	 \end{equation}
\end{subequations}
where
\begin{align}
	\mu_{\s}(\rho) = 
	(1- \rho)
	\left[
	1 
	- \alpha \gamma_{\s,\os} \rho
		\right] , \qquad \gamma_{\s,\s'} = \frac{2D_{\s}}{D_\s+D_{\s'}}  ,
		\label{equ:d-mix}
\end{align}
$\s, \s' \in \{ r, b\}$ \textcolor{black}{and $\os$  represents the opposite species to $\s$. } 
The constant  $\alpha$ is defined by the relationship
\beq
 \frac{\alpha}{1+\alpha} = \frac{1}{(2\pi)^d} \int_{[-\pi,\pi]^d} \frac{\sin^2\zeta_1}{  2\sum_k \sin^2 (\zeta_k/2)}  d\zeta. 
 \label{equ:alpha}
\eeq
\textcolor{black}{Note that, although not immediately apparent in its form above, $M^\text{low}$ is symmetric
\begin{equation}
	M^\text{low} = (1-\rho) \begin{pmatrix}
	D_r\rho_r & 0 \\
0 & D_b  \rho_b 
\end{pmatrix}  
- \frac{2 \alpha (1-\rho) \rho_r \rho_b}{D_r + D_b}\begin{pmatrix}
	D_r^2 & -D_r D_b\\
	-D_r D_b & D_b^2
\end{pmatrix}.
\end{equation}
}
The value of $\alpha$ depends on the dimension $d$ of the underlying lattice. One obtains
\beq
\alpha = (\pi/2) -1 \quad \text{for} \quad d=2,
 \label{equ:alpha-d2}
\eeq
while $\alpha \approx 0.265569$ for $d=3$. 
From a physical perspective, the coefficient $\alpha$ accounts for the fact that a particle diffusing in a particular direction tends to acquire an excess of particles in front of it, which acts to slow down its self-diffusion. A similar effect is observed in the continuous counterpart, namely Brownian hard spheres (see \cite{Maria2} or Eq. (7.9) in \cite{batchelor1976brownian}).

It is instructive to contrast \eqref{equ:M-mix} with the result one obtains after a simple mean-field closure of the equation of motion \cite{mean1} (see Subsect. \ref{sec:mean}), which fails to capture some cross-diffusion terms arising from the non-diagonal terms of the mobility matrix,
\begin{equation}
	M^{\text{mf}}(\rho_r, \rho_b) = (1-\rho) \begin{pmatrix}
	D_r\rho_r & 0 \\
0 & D_b  \rho_b 
\end{pmatrix}   .
	\label{equ:M-mf}
\end{equation}
This corresponds to \eqref{equ:M-mix} with $\mu_\s(\rho)$ replaced by $\mu_{\s}^\text{mf}(\rho) = 1- \rho$.
In Section \ref{sec:numerics} we show numerical results comparing the matched asymptotics and mean-field models, showing that the former significantly outperforms the latter.

\textcolor{black}{
%\paragraph{Global existence of weak solutions of the cross-diffusion system}
The existence of solutions of a PDE system describing macroscopic behaviour such as \eqref{equ:gen-hydro} is guaranteed when it is obtained rigorously as the hydrodynamic limit of a well-posed microscopic process \cite{singleasep,Quastel,Varbook,gabrielli1999onsager,seo2018scaling}. Conversely, when the PDE is obtained formally via a mean-field approximation or the method of matched asymptotics, existence must be established by other means. The mean-field cross-diffusion system (system \eqref{full_mainresult} with the mobility matrix replaced by \eqref{equ:M-mf}) was analysed in \cite{Burger}. Their global-in-time existence of weak solutions relies on the positive definiteness of the mobility matrix $M^{\text{mf}}$ and exploits the convexity and structure of the energy $E$ \eqref{equ:Efm}, which yields suitable bounds for $\rho_\sigma$. The method has been coined the \emph{boundedness-by-entropy method} \cite{jungel2015boundedness} and has since been used to show the existence of a variety of cross-diffusion systems \cite{berendsen2017cross}. In \cite{Burger} they are also able to show the uniqueness of the solution for initial data close to equilibrium.
Our system \eqref{full_mainresult} has the same energy (which should ensure that $\rho_\sigma\ge 0$ and $\rho < 1$) and a symmetric mobility matrix $M^\text{low}$. The matrix is always positive definite for $d=3$, while for $d=2$ it is only positive definite if the two diffusion coefficients are not too dissimilar.\footnote{The exact condition for \eqref{equ:M-mix} to be semi-positive definite for $d=2$ is that $\pi-3<D_b/D_r<1/(\pi-3)$. In fact, it is possible to remove the condition on the ratio of diffusivities by setting $\mu_{\s}(\rho) = (1- \rho)\left[1 - \alpha (1-\rho) \gamma_{\s,\os} \rho \right]$. This only changes $O(\rho^3)$ terms, and therefore, the underlying asymptotic equation remains unchanged at $O(\rho^2)$.} Provided that one can derive suitable a priori estimates (hypothesis H2' in \cite{jungel2015boundedness}), we would therefore expect a similar existence result to \cite{Burger} for our system \eqref{full_mainresult}.  
To the best of our knowledge, the uniqueness of solutions of the cross-diffusion system \eqref{full_mainresult} in a general sense (far from equilibrium) remains a delicate topic and is still an open problem \cite{jungel2017cross}. This would still be the case if the system was obtained as a hydrodynamic limit. 
}

%Within this general framework, an important technical question is the existence and uniqueness of solutions of \eqref{equ:gen-hydro}. We do not discuss this question in detail, but we note that the boundedness-by-entropy method method has been applied to similar cross diffusion systems  \cite{jungel2015boundedness,Burger,berendsen2017cross}, to establish existence. This relies on a symmetric positive semi-definite mobility and convex free-energy (or entropy). The free-energy \eqref{equ:Etot} is convex, and contains a $(1-\rho) \log (1- \rho)$ term, critical for controlling the density around $\rho =1$. Any consistent hydrodynamic equation must have $M$ positive semi-definite, to ensure diffusive behaviour. The specific case $M=M^\text{low}$ is always symmetric and is positive semi-definite for low density, when it is applicable\footnote{\textcolor{black}{For $d \geq 3$ then, $\alpha<1/2$ and $M^\text{low}$ is positive semi-definite provided $\rho \leq 1$. For $d=2$, this is only true for $\gamma_{\s,\os} < \frac{1}{ \alpha}$. For general $\gamma_{\s,\os}$, to ensure positive semi-definiteness for $\rho \leq 1$ one can set $	\mu_{\s}(\rho) = (1- \rho)\left[1 - \alpha (1-\rho) \gamma_{\s,\os} \rho \right]$. This only changes $O(\rho^3)$ terms and therefore the underlying asymptotic equation is still consistent at $O(\rho^2)$.}}. Such factors may enable a proof of existence using the boundedness-by-entropy method. On the other hand, uniqueness of solutions remains open for equations of this type. 

\paragraph{Expansion of the self-diffusion coefficient for \texorpdfstring{$D_b = D_r$}{DreqDb}} 

Finally, returning to the case $D_r=D_b=1$, one sees that \eqref{equ:M-mix} would coincide with the exact result \eqref{equ:M-quas} if $\mu_{\s}(\rho) \equiv D_s(\rho)$.  Using the variational representation of $D_s$ and applying a recursive approach proposed in~\cite{lamdim}, in Section \ref{sec:self-diff} we obtain the following polynomial expansion
	\begin{equation}
	\label{equ:Dexpan linear}
		{D_s(\rho)} = 
		\begin{cases}      	 
      	1-(1+\alpha) \rho +O(\rho^2) & \text{as }\rho \rightarrow 0,\\
     	\frac{1}{2\alpha+1}(1-\rho)+O((1-\rho)^2) & \text{as }\rho \rightarrow 1.
    \end{cases}     
	\end{equation}
Therefore, comparing \eqref{equ:Dexpan linear} and \eqref{equ:d-mix}, one does indeed have that the matched asymptotics result $\mu_\s$ agrees with the rigorous behaviour of $D_s$ near $\rho=0$ to first order in $\rho$, as expected (since the asymptotic expansion is computed to that order)\footnote{A note added in proof: Note added in proof: After submitting this manuscript, we became aware of the work of Nakazato and Kitahara on the self-diffusion coefficient of the same process under consideration \cite{Nakazato1980}. Their rational approximation of $D_s$ also agrees with (18) up to first order at both $\rho =0$ and $\rho = 1$.}. By contrast, the mean-field result $\mu_\s^\text{mf}(\rho) = 1-\rho$ already gets the first-order term in $\rho$ wrong.
We show additionally in Appendix \ref{appendix:matched second} that extending the matched asymptotic analysis to second order in $\rho$ yields an improved formula for $\mu$ that, again, agrees with the expansion of $D_s(\rho)$ to that order. This suggests that the (formal) method of matched asymptotics recovers the correct asymptotic behaviour of the hydrodynamic equation for $\rho \ll 1$. Subsection \ref{sec:self-diff:same} further shows (for $D_r=D_b$) that a polynomial ansatz for $D_s(\rho)$ provides numerically accurate predictions for the behaviour of the one-particle density in the whole range $\rho \in [0, 1)$. 
 
%%%%%%%%%%%%%%%%%%%%%%%%%%%%%%%%%%%%
\section{Analysis by method of matched asymptotics} \label{sec:MAE}
%%%%%%%%%%%%%%%%%%%%%%%%%%%%%%%%%%%%
\label{sec:matched}

%%%%%%%%%%%%%%%%%%%%%%%%%%%%%%%%%%%%
\subsection{Preliminaries}
\label{sec:Prelim}
%%%%%%%%%%%%%%%%%%%%%%%%%%%%%%%%%%%%

As in \cite{Burger,Quastel,Erignoux,AOUP,vJump,Varbook}, we seek a low dimensional PDE to describe the system. Here we derive a set of coupled ODEs for the one-particle densities, obtained from summation over the master equation of the simple exclusion process. These can then be interpreted as discretising a PDE system defined in $\TT^d$. 

%%%%%%%%%%%%%%%%%%%%%%%%%%%%%%%%%%%%
\subsubsection{Equation of motion for the one-particle probability density}
%%%%%%%%%%%%%%%%%%%%%%%%%%%%%%%%%%%%

The model defined in Subsection \ref{sec:model} has generator $\mathcal{L}$ whose action on a generic function $f : \J \to \mathbb{R}$ is 
\begin{equation}
	\label{Generator}
	\mathcal{L} f (\eta)=
	\sum_{y,z \in \Omega} 
	c(\eta, y,z) 
	\big[ f(\eta^{y z})-  f(\eta) \big]   ,
\end{equation}
where $\eta^{y z}$ denotes the configuration after the states of sites $y$ and $z$ have been swapped, that is
\begin{equation}
    \eta^{y z}(x) = 
    \begin{cases}
          \eta(x) &\text{ if }x\neq y,z,   \\
          \eta(y) &\text{ if }x = z,   \\
          \eta(z) &\text{ if }x = y. 
    \end{cases}
\end{equation}
Now take $f(\eta) = \eta^\s(x)$ and use the evolution equation, $\frac{d}{dt} \EE[ f ] = \EE [\mathcal{L} f ]$, with (\ref{equ:latt-rho}) to obtain
\begin{align}
	\dot \Rho_\s (x,t) &= \sum_{y \in \Omega} \EE \big[  c (\eta_t, y,x)\eta^\s_t(y) - c (\eta_t, x,y) \eta^\s_t(x) \big],
\end{align}
where the dot denotes the time derivative. 
The only jump rates that contribute to this expectation are those involving the movement of $\s$ particles (because any jump of the other species between sites $x,y$ would require that there is no $\s$ particle on either site).
Then by (\ref{equ:lambda-lambda}):  
\begin{align}
\label{equ:high density start}
\begin{aligned}
	\dot \Rho_\s (x,t)
	 = & \sum_{ y \in \Omega} 
	\lambda_\s  (y,x) \PP [ \eta^\s_t(y) =1, \eta^{\os}_t(x)= \eta^\s_t(x) =0 ]
	\\
	&- \sum_{ y \in \Omega}  
	\lambda_\s  (x,y) \PP [ \eta^\s_t(x) =1, \eta^{\os}_t(y) = \eta^\s_t(y) =0 ]. 
	\end{aligned}
\end{align}
Consider the marginal probability densities for the position of a single particle and for two particles, described by the functions
\begin{alignat}{3}
		\rrr (x,t) &= \frac{\EE[\eta^\s_t(x)]}{N_\s  h^d}, \nonumber \\ 
\label{equ:pij}
	P_{\s,\os} (x,y,t) & = \frac{ \EE [\eta^\s_t(x) \eta_t^{\os} (y) ] }{N_\s N_{\os} h^{2d}}, && \textrm{for } x\ne y ,\\
\nonumber
	P_{\s,\s} (x,y,t) & = \frac{ \EE [\eta^\s_t(x) \eta_t^{\s}(y) ] }{(N_\s -1) N_{\s} h^{2d}}, & \quad & \textrm{for } x\ne y  ,     
\end{alignat}
and $P_{\s,\os} (x,x,t) = P_{\s,\s} (x,x,t) = 0$. The densities are normalised so that $P_\s(x,t),~ P_{\s, \s'}(x,y,t)$ converge (as $h\to0$) to probability densities in $\TT^d$ and $\TT^{2d}$ respectively.
Note that the volume density \eqref{equ:latt-rho} is related to the probability density as
\begin{equation}
\Rho_\s(x,t) = N_\s  h^d \rrr (x,t)    .
\label{equ:rho-and-p}
\end{equation}

The discrete density $P_\s$ is a natural object when considering matched asymptotics because it remains $O(1)$ with respect to both $h \ll 1$ and $Nh^d \ll 1$.
Combining \eqref{equ:pij} and \eqref{equ:high density start} implies: 
\begin{align}
\label{equ: int 1}
\begin{aligned}
		\dot P_\s (x,t)  = &\sum_{ y \in \Omega}  \lambda_\s  (y,x) \left[ P_\s (y) - h^{d}N_{\os} P_{\s, \os}(y,x) - h^{d}(N_\s -1) P_{\s, \s}(y,x) \right]
	\\
	&- \sum_{ y \in \Omega}  \lambda_\s  (x,y) \left[ P_\s (x) - h^{d}N_{\os} P_{\s, \os}(x,y) - h^{d}(N_\s -1) P_{\s, \s}(x,y) \right], 
\end{aligned}
\end{align}
where we have omitted the time variable on the right-hand side for ease of presentation.
As usual, this equation of motion for the one-particle (marginal) density involves the two-particle density, so the system is not closed. We could go back to \eqref{Generator1} and obtain equations for the two-particle densities, but these in turn would depend on the three-particle densities. This is analogous to the BBGKY hierarchy~\cite{BBGKY} that one obtains when integrating the Liouville equation for continuous processes. In the following, we present two approaches to close \eqref{equ: int 1}, namely the mean-field closure or the method of matched asymptotics to approximate the terms involving the two-particle densities.
	
To facilitate the following derivation, it is useful to consider the motion of a single particle, without any interactions. The single $\sigma$-particle model has a generator, denoted by ${\cal L}_\s$ with adjoint
\begin{equation}
	\label{Generator1}
	\mathcal{L}^*_\s  Q(x)= \sum_{y \in \Omega} \left[ \lambda_\s  (y,x) Q(y)- \lambda_\s  (x,y) Q(x) \right],   
\end{equation}
defined for generic functions $Q : \Omega \to \mathbb{R}$.
Combining \eqref{Generator1} and \eqref{equ: int 1}, one obtains
\begin{subequations}
\label{integrated_eq}	
\begin{equation}
	\label{equ:gen int eq}
	\dot P_{\s} (x,t)
	= \mathcal{L}^*_\s  P_\s (x,t) + \mathcal{E}^{\rm int}_\s (x,t)    ,
\end{equation}
where the interaction term $\mathcal{E}^{\rm int}_\s$ includes the effects of the exclusion process and is given by
\begin{equation}
\label{equ:inter}
	\mathcal{E}^{\rm int}_\s (x,t) = 
	~h^d (N_\s-1)\mathcal{E}_{\s,\s}(x,t)
	+
	h^d N_\os \mathcal{E}_{\s,\os}(x,t)  , 
\end{equation}
\end{subequations}
where $\mathcal{E}_{\s,\s}$ corresponds to the effect of a pairwise interaction between same-species particles, and $\mathcal{E}_{\s,\os}$ corresponds to the effect on the evolution of $P_\s$ of a pairwise interaction with a particle of the opposite species. Note that the prefactors in these two terms account for the number of such pairwise interactions present in the system. The pairwise interaction terms are given in terms of the two-particle probability densities, 
\begin{equation}
	\mathcal{E}_{\s,\s'}(x,t) = ~\sum_{y \in \HH}\left[ \lambda_\s  (x,y)P_{\s,\s'} (x,y,t)-\lambda_\s  (y,x)P_{\s,\s'} (y,x,t) \right] ,
	\label{equ:inter-explicit}
\end{equation}
where $\s, \s' \in \{ r, b\}$.
The task in the following will be to estimate the two-body interaction term $\mathcal{E}^{\rm int}$, to obtain a closed equation for $P_\s $.  We first discuss the mean-field approach in Subsection \ref{sec:mean} and then consider the method of matched asymptotic expansions in Subsection \ref{sec:Asymp}.

%%%%%%%%%%%%%%%%%%%%%%%%%%%%%%%%%%%%
\subsection{Mean-field approximation}
\label{sec:mean}
%%%%%%%%%%%%%%%%%%%%%%%%%%%%%%%%%%%%

A simple ad-hoc closure to the one-body equation is obtained by assuming that the occupancies of sites $x,y$ by species $\s,\s'$ are independent, that is, 
\begin{equation}
P_{\s,\s'}(x,y,t) = P_\s (x,t) P_{\s'}(y,t), \qquad x\neq y.
\label{equ:mf-closure}
\end{equation}
This is called the mean-field closure \cite{meanref1,meanref2,meanref3} .
Substituting this into \eqref{equ:inter-explicit} and using \eqref{equ:gen int eq} yields a closed system for $P_r, P_b$, namely
\begin{subequations}
	\label{mfa}
\begin{equation}
\dot P_\sigma(x,t) = \mathcal{L}^*_\s P_\sigma(x,t) + {\cal E}^{\rm mf}_\sigma(x,t),
\label{equ:mf-disc}
\end{equation}
with
\begin{align}
\begin{aligned}
	\mathcal{E}^{\rm mf}_\s &(x,t) =h^d N_\os \sum_{y \in \HH}\left[ \lambda_\s  (x,y) P_{\s} (x,t) P_{\os} (y,t)-\lambda_\s  (y,x)P_{\s}(y,t)P_{\os} (x,t) \right] \\ 
	&\quad +h^d (N_\s  -1) \sum_{y \in \HH}\left[ \lambda_\s  (x,y)P_{\s} (x,t)P_{\s} (y,t)-\lambda_\s  (y,x)P_{\s}(y,t)P_{\s} (x,t) \right], 
	\end{aligned}
\end{align}
for $\s \in \{ r, b\}$ and $\os$ the opposite species.
\end{subequations}

%%%%%%%%%%%%%%%%%%%%%%%%
\subsubsection{Cross-diffusion system in continuous space}
%%%%%%%%%%%%%%%%%%%%%%%%
\textcolor{black}{ 
In the next step we assume that the hydrodynamic limit exists, and that $P_\s$ converge to smooth densities $p_\s$ on $\TT^d$,
\begin{equation}
	\label{rel_P_p}
	P_\s(x,t) = p_\s(x,t) + O(h), \qquad \text{for } x\in \Omega.
\end{equation}
This is based on the physical intuition that small-scale fluctuations in the particle density are quickly smoothed out by rapid local mixing of the system. Therefore the discrete densities $P_\s(x,t)$ on $\Omega \subseteq (h\ZZ)^d$ should not vary rapidly across lattice sites.} We also expand the rate \eqref{equ:rates-lambda-i} for $\vert  x-y \vert  =h$ and small $h$: 
\beq
\label{equ:lambda_expan}
\lambda_\s(x,y) = \frac{D_\s}{h^2}+\frac{D_\s}{h^2}\frac{x-y}{2}\cdot\nabla V_\s \Big( \frac{y+x}{2} \Big)+O(1).
\eeq
Inserting \eqref{equ:lambda_expan} into \eqref{mfa} and letting $h \to 0$ while keeping the volume fractions $\phi_\sigma$ fixed, one finds
\begin{align}\label{limit_p_mfa}
\begin{aligned}
	\partial_t p_\sigma(x,t) &= D_\sigma \nabla \cdot \left[ \nabla p_\sigma + p_\sigma \nabla V_\sigma  \right] - \phi_\sigma D_\sigma \nabla \cdot \left[  p_\sigma^2 \nabla V_\sigma \right] \\
	& \quad  + \phi_{\os}  D_\sigma \nabla \cdot \left[ p_\sigma \nabla p_{\os} - p_{\os} \nabla p_\sigma - p_\sigma p_{\os} \nabla V_\sigma \right].
\end{aligned}
\end{align}
The first divergence on the right-hand side corresponds to linear drift-diffusion of a free $\sigma$-particle, and the second and third terms (premultiplied by $\phi_\sigma$ and $\phi_{\os}$ respectively) arise from the simple exclusion rule. 
Multiplying \eqref{limit_p_mfa} by $\phi_\s$, the average particle density $\rho_\s$ [the continuous analogue of \eqref{equ:rho-and-p}] solves the hydrodynamic PDE
\begin{equation}
\label{equ:2body}
	\p_t \rho_\sigma = D_\s \nabla \cdot \Big[ (1-\rho) \nabla \rho_\s + \rho_\s \nabla \rho + (1-\rho) \rho_\s \nabla V_\s  \Big]   ,
\end{equation}
which was obtained previously in \cite{Burger,mean1}. The system of equations \eqref{equ:2body} for $\rho_r, \rho_b$ can be written in gradient-flow form \eqref{equ:gen-hydro} with mobility \eqref{equ:M-mf} and energy \eqref{equ:Etot}.

%%%%%%%%%%%%%%%%%%%%%%%%%%%%%%%%%%%%
\subsection{Low-density approximation via matched asymptotics}
\label{sec:Asymp}
%%%%%%%%%%%%%%%%%%%%%%%%%%%%%%%%%%%%

Deviations from the ad-hoc mean-field closure \eqref{equ:mf-closure} are due to particle correlations, which are maximised when particles occupy adjacent sites.  Since the interaction term $\mathcal{E}^{\rm int}_\s$ in \eqref{equ:inter} is evaluated exactly at these configurations, one may expect significant correlation effects. Here we approximate the interaction term using the method of matched asymptotics. This is a systematic asymptotic method, well-suited to study problems with boundary layers governed by a small parameter, and previously used to study systems of interacting Brownian hard spheres \cite{Maria1,Maria2}. 
In contrast to the mean-field approach, this systematic procedure does not require a closure assumption and leads to a controlled approximation of $\mathcal{E}^{\rm int}_\s$ as an asymptotic series in $\phi$ and $h$. 
Here we adopt this procedure for the discrete simple exclusion process: consistently with the hydrodynamic limit, the asymptotic expansions at small $h$ take place at fixed $\phi_\s = N_\s h^d$. But additionally (in this Section), we assume that the total occupied fraction $\phi = Nh^d = \phi_r + \phi_b$ is also small, such that we can asymptotically expand the equation for the two-particle density in powers of $\phi$.  

The interaction term \eqref{equ:inter-explicit} depends on the two-particle density $P_{\sigma, \sigma'}$, so in order to approximate it we consider the evolution equation of $P_{\sigma, \sigma'}$. To this end, we set $f(\eta) = \eta^\s(x) \eta^{\s'}(y)$ in \eqref{Generator} and follow a similar calculation to Section \ref{sec:Prelim} to obtain (expanding asymptotically in $\phi$)
\begin{multline} \label{equ:two particle}
	\dot P_{\s,\s'}(x,y,t) =
	\sum_{ \substack{z \in \HH \\ z \neq y} }
	 \left[
	\lambda_{\s}  (z,x) P_{\s,\s'}(z,y,t) -\lambda_{\s}  (x,z) P_{\s,\s'}(x,y,t)
	\right]
\\
	+ \sum_{ \substack{z \in \HH \\ z \neq x} }
	 \left[
	\lambda_{\s'}  (z,y) P_{\s,\s'}(x,z,t) -\lambda_{\s'}  (y,z) P_{\s,\s'}(x,y,t)
	\right]
	+O(\phi)  ,
\end{multline}
where the higher-order terms contain particle interactions. This is because, when particles are dilute ($\phi \ll 1$), the probability of configurations with three or more particles nearby is much less than that of configurations with only two particles close by.  
We exploit the fact that, to leading-order in $\phi$, \eqref{equ:two particle} is a closed equation for $P_{\sigma, \os}$, so that we can effectively focus on the $N=2$ problem, while deriving a consistent approximation of ${\cal E}_{\rm int}$. 
Extension to higher orders in $\phi$ is discussed in Appendix~\ref{appendix:matched second}.

%%%%%%%%%%%%%%%%%%%%$$$
\subsubsection{Inner and outer regions} \label{sec:inout}
%%%%%%%%%%%%%%%%%%%%%%%

At leading order, \eqref{equ:two particle} can be interpreted as a master equation for a two particle system (of types $\sigma,\sigma'$) where $(x,y) \in \Omega^2 \setminus \{x=y\}$ are the particle positions. We expect the solution to \eqref{equ:two particle} to display a boundary layer near the excluded diagonal $x=y$ due to strong correlations arising from the simple exclusion rule. These correlations decay as the separation distance $\vert x-y\vert$ grows. 
 This motivates the use of matched asymptotic expansions, with an \emph{outer region} in which the two particles are well-separated ($\vert x-y\vert  \gg h$) and an \emph{inner region} in which the particles are close to each other ($\vert x-y\vert  \sim h$). Here, in contrast to the standard approach~\cite{Maria1,Maria2}, the inner or boundary layer variable will be discrete, and only the outer variable will be continuous. This enables an accurate characterisation of $P_{\s,\s'}$ as $h\to 0$; 
in particular it allows us to keep the exact geometry of the interaction between two particles and accurately evaluate the interaction term ${\cal E}_{\rm int}$. With this in mind, we assume that there exist functions $P_{\rm out} : \TT^2 \times \RR_+ \to \RR$ and $P_{\rm in} : \TT \times \ZZ^d \times \RR_+ \to \RR$ such that $P_{\s,\s'}$ can be written for small $h$ as
\beq
P_{\s,\s'}(x,y,t) = \begin{cases} P_{\rm out}(x,y,t), & \vert x-y\vert  \gg h, \\
P_{\rm in}\big(x,\frac{y-x}{h},t\big), & \vert x-y\vert  \sim h,
\end{cases}
\label{equ:out-in}
\eeq
where the dependence of $P_{\rm out}, P_{\rm in},$ on $\s, \s'$ is left implicit for compactness of notation. We establish asymptotic approximations for $P_{\rm out}$ and $P_{\rm in}$, valid in the outer and inner regions respectively.  We enforce that they agree in the crossover between the two regions by imposing a \emph{matching condition} as explained below.

By assumption, the size-exclusion rule appearing as the condition $z \ne y$ or $z \ne x$ in \eqref{equ:two particle} does not appear in the outer region. Recalling $P_{\sigma,\sigma'} (x,y,t) = P_{\rm out}(x,y,t)$ in the outer region, we obtain
\begin{align}
\label{equ:two outer}
\begin{aligned}
	\dot P_{\text{out}}(x,y,t) &=
	\sum_{ \substack{z \in \HH } }
	 \left[
	\lambda_{\s}  (z,x) P_{\text{out}}(z,y,t) -\lambda_{\s}  (x,z) P_{\text{out}}(x,y,t)
	\right]
\\
	& \phantom{=} + \sum_{ \substack{z \in \HH} }
	 \left[
	\lambda_{\s'}  (z,y) P_{\text{out}}(x,z,t) -\lambda_{\s'}  (y,z) P_{\text{out}}(x,y,t)
	\right]
+O(\phi) \\
&=  \mathcal{L}^*_\s P_{\text{out}}(\cdot,y,t) + \mathcal{L}^*_{\s'} P_{\text{out}}(x,\cdot,t) +O(\phi)  ,
\end{aligned}
\end{align}
using the independent walk or single particle adjoint generator \eqref{Generator1} (we write $\mathcal{L}^*_\s P_{\text{out}}(\cdot,y,t)$ to denote the operator acting on $P_{\text{out}}(x,y,t)$ as if it was a function of $x$ only for $y$ and $t$ fixed). Therefore, at leading order in $\phi$, the evolution of $P_{\text{out}}$ corresponds to two independent $\sigma$ and $\sigma'$ particles. That is, $P_{\text{out}}(x,y,t) = Q_\sigma(x,t) Q_{\sigma'}(y,t)+O(\phi)$ for some functions $Q_\sigma$ satisfying $\dot Q_\s = \mathcal{L}^*_\s Q_\sigma$.
Using the normalisation condition, $ \sum_{x,y} h^{2d} P_{\s,\s'} = 1$, and \eqref{rel_P_p}
\begin{equation}
\label{equ:outer lead}
	P_{\text{out}}^{} (x,y,t) = p_{\s} (x,t)p_{\s'} (y,t) + hP_{\text{out}}^{(1)}(x,y,t)+ O(h^2,\phi)  .
\end{equation}
The correction term at $O(h)$ comes from approximating discrete densities by continuous densities for $h \ll 1$. 
As expected, particles are independent to leading order in the outer region (so the outer solution `does not see' the interaction rule). 

In the inner region ($\vert x-y\vert  \sim h$), we introduce \emph{inner variables} $\u ,  \v$ satisfying $x = \hat x$ and $y = \hat x + h \v$. Then the \emph{inner density} $P_{\rm in}(\u, \v,t)$ is equal to $P_{\s, \s'}(x,y,t)$, by \eqref{equ:out-in}. As discussed above, the inner density is taken to be continuous with respect to its first argument $\hat x$, while keeping its second argument $\v\in \ZZ^d$  discrete, to parameterise the boundary layer.

\begin{figure}
\centering
	\includegraphics[width=.5\textwidth]{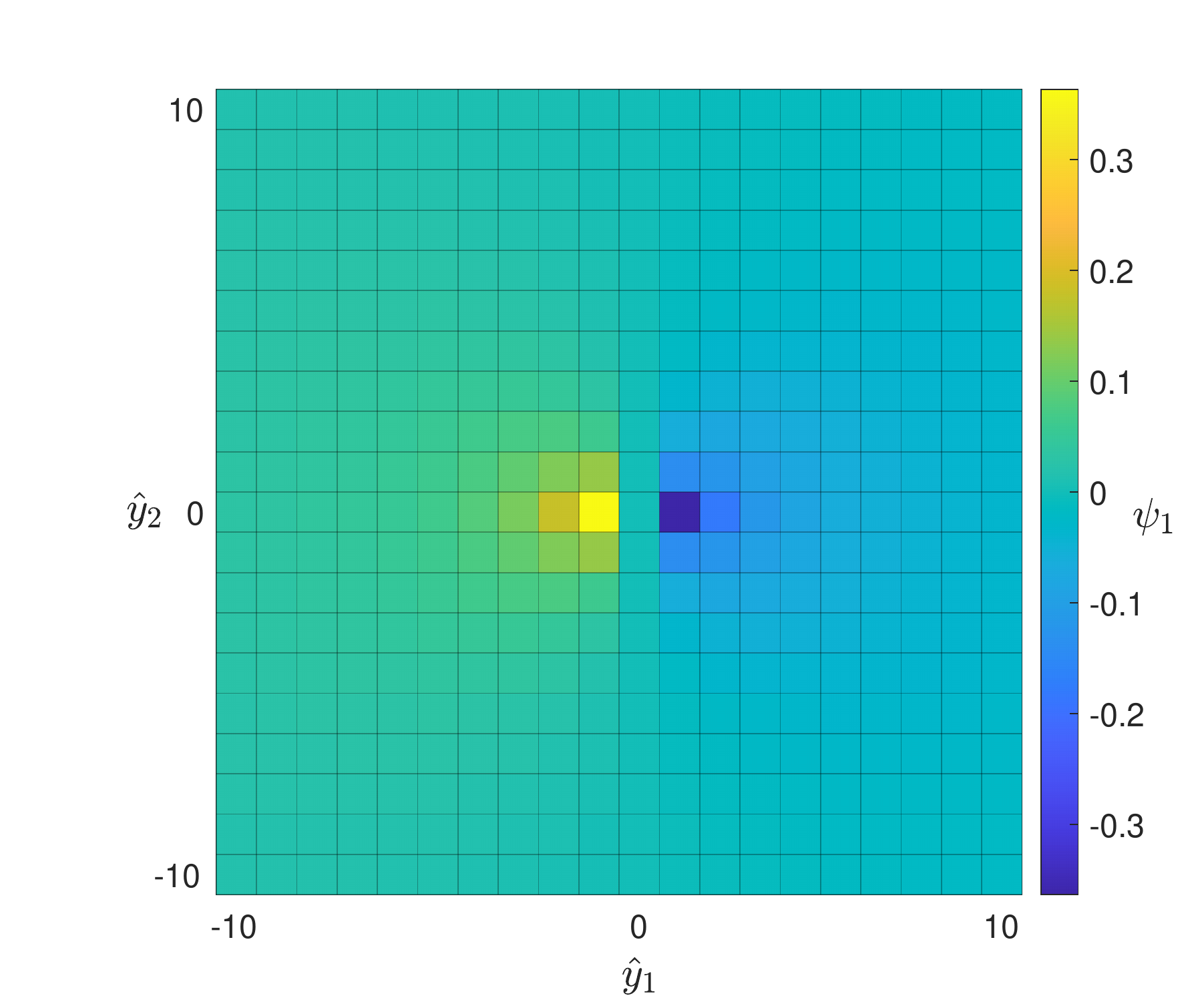}
	\caption{Solution of the auxiliary function $\psi_1$ satisfying \eqref{aux_problem}, shown for $d=2$ and $\hat y \in \{-10,-9, \dots, 10\}^2$, and evaluated using \eqref{equ:psi-integral}.}
	\label{fig:greens}
\end{figure} 

To understand the qualitative dependence of $P_{\rm in}$ on its second argument $\hat y$, consider Fig.~\ref{fig:greens}, which displays the auxiliary function $\psi_1$ that solves (\ref{aux_problem}), below. (It will be shown that the dependence of $P_{\rm in}$ on $\hat y$ is similar to that of this function.) The correlation between particles, neglected by the mean-field approximation, is captured at first order by $\bgfpsi$. 
For $\hat y$ close to the origin, $\psi_1(\hat y)$ differs significantly between adjacent sites, it describes a boundary layer that is directly affected by the structure of the lattice.  For larger $\hat{y}$, the function decays in modulus, and the relative differences between adjacent sites also decay.  This latter property is required because $P_{\rm in}$ must obey a matching condition with a function $P_{\rm out}$ that depends smoothly on its second argument, as $\vert\hat{y}\vert\to\infty$, recall (\ref{equ:out-in}).

We note that, in the inner region, the summations over the auxiliary variable $z$ in \eqref{equ:two particle} only lead to nonzero terms if $z$ is at a distance $h$ from either $x$ or $y$. Therefore, introducing the set $B_\v = \{ e \in \ZZ^d : \vert e \vert  =1, e \neq \v \}$, the first summation $\{z \in \HH, z \neq y\}$ reduces to $z = \u + h \z$ and the second summation $\{z \in \HH, z \neq x\}$ to $z = \u + h(\v-e)$, with $\z \in B_\v $. Changing to inner variables in \eqref{equ:two particle} gives, to order $\phi$,
\begin{subequations}
	\label{inner_problem}
\begin{align}
	\dot P_{\text{in}}(\u,\v,t) &= \! \! \sum_{ \z \in B_\v } 
	 \! \left[
	\lambda_{\s}  (\u+h\z,\u) P_{\text{in}} (\u+h\z,\v-\z,t) -\lambda_{\s}  (\u,\u+h\z) P_{\text{in}} (\u,\v,t)\right] \nonumber \\
	& \phantom{=} + \! \sum_{ \z \in B_\v}  \!
	 \big[\lambda_{\s'}  (\u + h(\v-\z),\u + h\v) P_{\text{in}} (\u,\v-\z,t) \nonumber \\
	 & \phantom{= + \sum_{ \z \in B_\v} 
	 \big[} - \lambda_{\s'}  (\u + h\v, \u + h(\v-\z)) P_{\text{in}} (\u,\v,t) \big], \label{equ:two inner}
\end{align}
for $\u\in \Omega \subseteq (h\ZZ)^d$ and $\v \in \zdstar$ with $\zdstar = \ZZ^d \setminus \{ 0\}$.
In the first line above, we have used that $P_{\text{in}} (\u+h\z,\v-\z,t) = P_{\s,\s'}(\u+h\z, \u+h\z + h(\v-\z), t)$, which is equal to the desired $P_{\s,\s'}(z, y, t)$ converting back to the original variables. 

The inner solution $P_{\text{in}}$ must match with the outer solution $P_{\text{out}}$ as $\vert \v \vert \to \infty$. Writing $P_{\text{out}}$ in \eqref{equ:outer lead} in terms of the inner variables and expanding gives
\begin{align} \label{matching}
P_{\text{out}}^{} &\sim  p_{\s} (\u,t)p_{\s'} (\u + h \v,t) + hP_{\text{out}}^{(1)}(\u ,\u + h \v,t)\\
						& \sim p_{\s} (\u,t)p_{\s'} (\u,t) + h p_{\s} (\u,t)   \v \cdot \nabla p_{\s'} (\u,t)+ hP_{\text{out}}^{(1)}(\u ,\u ,t) ~~~ \textrm{as } \vert \v \vert \to \infty. \nonumber
\end{align}
\end{subequations}
Next, we seek a solution to the inner problem \eqref{inner_problem} of the form
$$
P_{\text{in}} (\u,\v,t) = P_{\text{in}}^{(0)} (\u,\v,t) +h P_{\text{in}}^{(1)} (\u,\v,t) + \cdots.
$$
Using \eqref{equ:lambda_expan} and expanding $P_{\text{in}}$ with respect to its first argument (which we recall is continuous), the leading-order inner problem (which comes at  order $h^{-2}$) is
\begin{subequations}
	\label{inner0}
\begin{equation} \label{inner0eq}
	0 =
	\sum_{ e \in B_\v } %\substack{\vert e\vert  =1 \\ e \neq v} }
	 ( D_\s+D_{\s'} )\left[
	 P_{\text{in}}^{(0)} (\u, \v-\z,t) - P_{\text{in}}^{(0)} (\u, \v,t) 
	\right] , \qquad \v \in \zdstar,
\end{equation}
together with the condition at infinity
\begin{equation} \label{inner0match}
	P_{\text{in}}^{(0)} (\u,\v,t) \sim p_{\s} (\u,t) p_{\s'} (\u,t)  \qquad \text{as} \qquad \vert\v\vert \rightarrow \infty   .
\end{equation}
\end{subequations}
It is straightforward to see that a function constant in $\v$ satisfies \eqref{inner0eq}. Thus, using \eqref{inner0match} we find that the (trivial) solution for the inner problem \eqref{inner0} is
\begin{align}
\label{equ:inner lead}
	P_{\text{in}}^{(0)} (\u,\v,t) = p_{\s} (\u,t) p_{\s'} (\u,t)     .
\end{align}
In what follows we simply write $p_\s$ for $p_\s (\u,t)$. 
At $O(h^{-1})$, \eqref{inner_problem} reads, using \eqref{equ:inner lead} and \eqref{equ:lambda_expan}, 
\begin{subequations}
	\label{inner1}
\beq
\label{equ:inner first problem}
0 = \sum_{ e\in B_\v} 
	\left[
	 P_{\text{in}}^{(1)} (\u, \v-\z,t) - P_{\text{in}}^{(1)} (\u, \v,t)
	\right]  -  \v \cdot \bfkappa(\u,t)   \id_{\{ \vert \v\vert =1  \} }, \quad \v \in \zdstar,
\eeq
with
\beq
\bfkappa(\u,t) = \frac{1}{ D_\s+D_{\s'}} \left[D_\s \nabla \left( p_{\s} p_{\s'}  \right) 
 + (D_\s \nabla V_\s - D_{\s'} \nabla V_{\s'} ) p_{\s} p_{\s'}\right],
\eeq
together with the matching condition 
\begin{equation}
	P_{\text{in}}^{(1)} (\u,\v,t) \sim  \bfB(\u,t) \cdot \v + \bfC(\u,t) \qquad \textrm{as } \vert \v \vert \to \infty  ,
\end{equation}
where $\bfB(\u, t) = p_{\s}(\u, t) \nabla p_{\s'}(\u, t)$, $\bfC( \u ,t) = P_{\text{out}}^{(1)}(\u ,\u ,t)$. 
\end{subequations} 
In order to solve \eqref{inner1}, we first define the auxiliary problem 
\begin{subequations} \label{aux_problem}
	\begin{alignat}{3}
		 \lap \psi_j  &=   \v_j   \ind ,& \quad & \v \in \zdstar,\\
	\psi_j &\sim 0, & &\vert \v\vert \to \infty. 
	\end{alignat}
\end{subequations}
where $\lap$ is the standard discrete Laplacian (with unit grid spacing).
We remark that for $d \geq 3$, the auxiliary function $\psi_j$ can be related to the discrete Green's function. 
Its properties are discussed in Appendix~\ref{app:psi}, see also Fig.~\ref{fig:greens}.  In particular
\begin{equation} \label{prop_green}
	\psi_j (0) = 0, \qquad \psi_j(\pm e_k) = \pm \beta \delta_{jk},
\end{equation}
where 
\begin{equation}
	\label{beta_def}
	\beta = -\frac{1}{(2\pi)^d} \int_{[-\pi,\pi]^d} \frac{\sin^2 \zeta_1}{ 2\sum_k \sin^2 (\zeta_k/2)}  d\zeta ,
\end{equation}
is a $d$-dependent constant with $\beta\in(-1,0)$ for $d\geq 2$.
It is related to the constant $\alpha$ of (\ref{equ:alpha},\ref{equ:alpha-d2}) as
\begin{equation}
	\label{alpha_def}
	\alpha = -\frac{\beta}{\beta+1} \textcolor{black}{\in (0,\infty)}.
\end{equation}
Using (\ref{aux_problem},\ref{prop_green}), we find that
\begin{align}
		\sum_{ e\in B_\v} 
	\left[
	 \psi_j (\v-\z) - \psi_j (\v) 
	\right]  &  = \lap \psi_j + \left[ \psi_j(\v) - \psi_j(0)\right] \ind 
	\nonumber \\ & = 
	 (1+ \beta) \v_j \ind. 
\end{align}
Therefore, the solution to \eqref{inner1} can be identified as
\begin{align}
	\label{equ:inner first2}
		P_{\text{in}}^{(1)} & = \frac{1}{1+\beta}(\bfkappa - \bfB) \cdot {\bgfpsi}(\v) + \bfB \cdot \v +\bfC
			\nonumber \\ 
& = \bfA \cdot {\bgfpsi}(\v) + \bfB \cdot \v +\bfC,
\end{align}
where
\begin{align*}
\bfA(\u,t) = \frac{1}{1+\beta} \bigg[& \left( \frac{D_\s}{D_\s +D_{\s'}} \right) p_{\s'} \left( \nabla p_\s + p_\s \nabla V_\s \right) \\
& - \left( \frac{D_{\s'}}{D_\s +D_{\s'}} \right) p_{\s}  \left( \nabla p_{\s'} +  p_{\s'}\nabla V_{\s'} \right) \bigg].
\end{align*}
Hence, combining \eqref{equ:inner lead} and \eqref{equ:inner first2} we arrive at
\begin{align}
	\label{equ:inner soln} 
	P_{\text{in}} (\u , \v, t)  = 
		p_\s (\u , t) p_{\s'} (\u , t) + h \left[ \bfA \cdot {\bgfpsi}(\v) + \bfB \cdot \v +\bfC \right]+O(h^2, \phi).
\end{align}

%%%%%%%%%%%%%%%%%%%%%%%%%%%%%%%%%%%
\subsubsection{Systems of equations for \texorpdfstring{$P_\s$}{Psigma}}
%%%%%%%%%%%%%%%%%%%%%%%%%%%%%%%%%%%

The final step of this computation is to evaluate the interaction terms $\mathcal{E}_{\s,\s'}$ in (\ref{equ:inter-explicit}), to obtain a closed set of equations for $P_\s$.
The summands in (\ref{equ:inter-explicit}) are zero unless $y$ is adjacent to $x$, so in particular, they are zero in the outer region. Therefore we use the inner solution $P_{\rm in}$ to evaluate them.  
In inner variables \eqref{equ:inter-explicit} reads,
\begin{multline}
	\mathcal{E}_{\s,\s'}(x,t) = 
\frac{D_\s}{h^2} \sum_{\substack{\hat y \in \ZZ^d \\ \vert  \hat y \vert = 1}}\left\{1 +\frac{h^2}{8} [\hat y \cdot\nabla V_\s (\hat x) ]^2\right\} \left[P_{\rm in}(x , \hat y) - P_{\rm in}(x + h \hat y , -\hat y) \right] 
\\
-\frac{D_\s}{2h} \sum_{\substack{\hat y \in \ZZ^d \\ \vert  \hat y \vert = 1}}\left[ \hat y \cdot\nabla V_\s (\hat x) + \frac{h}{2} \hat y\cdot D^2 V_\s(\hat x) \hat y \right] \left[ P_{\rm in}(x , \hat y) + P_{\rm in}(x + h \hat y , -\hat y) \right]\\
	   +O(h, \phi) 
	 .
\end{multline} 
Combining with \eqref{equ:inner soln} yields: 
\begin{equation}
	\mathcal{E}_{\s,\s'} = 
D_\s \nabla \cdot \left[ 2\beta \bfA + 2\bfB - \nabla( p_\s p_{\s'}) - \nabla V_\s p_\s p_{\s'} \right]
	   +O(h, \phi)
	 .
\end{equation}
By evaluating $\bfA$ and $\bfB$ and taking the limit $h \to 0$ it follows that
\begin{multline}
		\label{equ:inter asymp}
	\mathcal{E}_{\s,\s'} = D_\s \nabla \cdot \Big[ 
	(1+ \alpha \gamma_{\s',\s})p_\s \nabla p_{\s'} 
	- (1+ \alpha \gamma_{\s,\s'} ) p_{\s'} \nabla p_\s 
	\\
	- (1+ \alpha \gamma_{\s,\s'} ) p_\s p_{\s'}\nabla V_\s 
	+ \alpha \gamma_{\s',\s} p_\s p_{\s'}\nabla V_{\s'}
	\Big]    +O(\phi)
	 ,
\end{multline}
with
\beq \label{gamma_def}
\gamma_{\s,\s'} = \frac{2D_{\s}}{D_\s+D_{\s'}},
\eeq
noting that $\gamma_{\s,\s} = 1$. 
Finally, taking the limit $h\to 0$ in \eqref{integrated_eq} yields
\begin{equation}
		\label{equ:asymp soln}
\partial_t p_\sigma(x,t) = D_\s \nabla \cdot \left[ \nabla p_\sigma + p_\sigma \nabla V_\sigma  
\right]  
+\phi_\s {\cal E}_{\s,\s} + \phi_\os{\cal E}_{\s, \os}
+ O(\phi^2)   .
\end{equation}
where the ${\cal E}$ factors are given by \eqref{equ:inter asymp}, resulting in a closed set of equations for the $p_\sigma$.

Note that we have focused throughout on the limit where $h\to0$ and $N_\sigma\to\infty$ at fixed $\phi_\sigma$, but \eqref{equ:asymp soln}
is still valid if $N_\s=1$ for some species $\s$.  In this case $\phi_\s=h^d$ tends to zero in the limit, and the terms corresponding to interactions with this species vanish in~\eqref{equ:asymp soln}.  (Physically, the single particle of species $\s$ has a negligible effect on the collective motion of the other particles in the system.)

Multiplying \eqref{equ:asymp soln} by $\phi_\s$, the average particle density $\rho_\s$ [the continuous analogue of \eqref{equ:rho-and-p}] solves the hydrodynamic PDE
\begin{multline}
\label{eq:mainresult}
		\p_t \rho_\sigma = D_\s \nabla \cdot \Big[ (1-\rho) \nabla \rho_\sigma + \rho_\sigma \nabla \rho + (1-\rho) \rho_\s \nabla V_\s
		\\
		- \alpha \gamma_{\s,\os} \rho_{\os} ( \nabla \rho_\sigma +\rho_\sigma \nabla V_\s) + \alpha \gamma_{\os,\s} \rho_\sigma( \nabla \rho_{\os} + \rho_{\os} \nabla V_{\os}) \Big] + O(\rho^3)   .
\end{multline}
Upon factorisation this equation takes the form \eqref{equ:gen-hydro}: it is a gradient flow with energy \eqref{equ:Etot} and mobility \eqref{equ:M-mix}, as anticipated in Section \ref{sec:model}.  This is the main result of the matched asymptotic computation.

%%%%%%%%%%%%%%%%%%%%%%%%%%%%%%%%%%%%
\section{Self-diffusion coefficient: asymptotics and connection to rigorous results}
\label{sec:self-diff}
%%%%%%%%%%%%%%%%%%%%%%%%%%%%%%%%%%%%

The self-diffusion coefficient measures the effective diffusion coefficient of a \emph{single} tagged particle in an interacting environment with many other particles. 
The standard definition of a self-diffusion coefficient assumes that the environment is in a homogeneous equilibrium state.  In our context, this corresponds to setting $V_\sigma \equiv 0$ in the microscopic hopping rates \eqref{equ:rates-lambda-i} and stationary densities $\Rho_\s(x, t) \equiv \phi_\s$ (so that particles evolve according to a SSEP), where we recall that $\phi_\s = N_\s h^d$ is the fraction of sites occupied by particles of the $\s$-species. 

Because the self-diffusion coefficient is a macroscopic property of an individual particle, its analysis in the current framework requires tagging a single particle in the population-level model.
Here we see one big advantage of the matched asymptotic approach: Eq.~\eqref{equ:asymp soln} is still applicable if we take \textcolor{black}{$\s$ to be the `species' of a tagged particle}; then the evolution equation for this species yields the self-diffusion constant.
Usually, all the particles in the environment are taken to be identical (same species) \cite{lamdim,sphondiff}, and the tagged particle is simply a `coloured' particle, leading to a self-diffusion coefficient depending on the occupied fraction $\phi$. However, within our framework, it is possible to consider an environment consisting of a mixture of particles (say red and blue particles), leading to a self-diffusion coefficient depending on $\boldsymbol{\phi} = (  \phi_r, \phi_b)$.  To this end, we consider the system \eqref{equ:asymp soln} with three species $\s = \{r,b, g\}$ corresponding to red, blue, and green particles, where $N_g = 1$ so that the green species is the tagged particle.

The self-diffusion coefficient of a $\s$-particle is given by the limit\footnote{In the case $D_r = D_b$ the limit is proven to exist \cite{kipnis1986}. In the case $D_r \neq D_b$ formally we can infer the limit exists via the limiting  PDE \eqref{equ:pdeDsneqD}. }
\begin{equation}
\label{equ:self-diff}
D_{s,\s} (\boldsymbol{\phi} ) =  \lim_{t \rightarrow \infty} \frac{1}{t} \EE \left( \frac{\vert  X_t - X_0 \vert ^2 }{2d} \right),
\end{equation}
where $X_t \in (h \ZZ)^d$ for $d>1$ denotes the position of the tagged particle at time $t$. The physical effect that controls $D_{s,\s}$ is that when the tagged particle makes a hop in a given direction, it leaves an empty site behind it. For $\phi>0$, this means that the particle's next jump is more likely to return to its original location, compared to other adjacent sites.  Over many jumps, this generates a `density wave'  in front of the tagged particle, which tends to suppress further motion in the same direction.  Hence one expects $D_{s,\s}<D_\s$.
(In one dimension, this is effect is so strong that the tagged particle is subdiffusive, $D_{s}=0$.)

Note also that the definition \eqref{equ:self-diff} applies to the model defined on the infinite lattice, $(h\ZZ)^d$, and in this case, the scaling of the hopping rates $\lambda_\s$ with $h$ means that the right-hand side of \eqref{equ:self-diff} is independent of $h$.  
On the other hand, to estimate $D_{s,\s}$ using a periodic lattice, we approximate this same quantity as $D^h_{s,\s}$, by taking $X_t - X_0$ as the relative displacement (i.e., the sum of jumps taken up until time $t$). We expect that $D^h_{s,\s} \rightarrow D_{s,\s}$ as $h \to 0$ \footnote{To understand how $D^h_{s,\s}$ differs from $D_{s,\s}$, note that the density wave that forms in front of the diffusing particle can loop round the periodic boundaries and interact with the other side of the particle.  This tends to cause $D^h_{s,\s}>D_{s,\s}$. However, for $h \ll 1$ (i.e $L = 1/h \gg 1$) this difference will be small. (For a single species, it is proven that $D^h_{s,\s} \rightarrow D_{s,\s}$ \cite{finitelatticediff}.)}.

%%%%%%%%%%%%%%%%%%%%%%%%%%%%%%%%%%%%
\subsection{Self-diffusion for \texorpdfstring{$D_r  \neq D_b$}{DrneqDb} } 
\label{sec:self-diff:Diff}
%%%%%%%%%%%%%%%%%%%%%%%%%%%%%%%%%%%%

The result (\ref{equ:asymp soln}) can be used to compute the self-diffusion coefficient for small $\phi$. As described above, we consider three species and take the tagged particle to be the only member of the green species $\ta$. Setting $V_\s\equiv 0$ (for all species) and $N_\ta = 1$, the generalised form of  \eqref{equ:asymp soln} is
\begin{equation}
 \partial_t p_\ta = D_\ta \nabla^2 p_\ta  
+  \sum_{\s' \in \{r,b\}} \phi_{\s}{\cal E}_{g,\s} 
+ O(\phi^2)  .
\end{equation}
where ${\cal E}_{g,\s}$ is given by \eqref{equ:inter asymp}.  
Combining with that equation yields
\begin{multline} \label{equ:pdeDsneqD}
    	\p_t p_\ta = D_\ta \nabla \cdot \Bigg\{ \bigg[1- \sum_{\s \in \{r,b\}} \nolimits (1+\alpha \gamma_{\ta,\s}) \rho_{\s}\bigg]\nabla p_\ta  \\
    	+ p_\ta \sum_{\s \in \{r,b\}} \nolimits (1+\alpha \gamma_{\s,\ta}) \nabla \rho_{\s}
\Bigg\}
+ O(\phi^2) ,
\end{multline}
where $\rho_\s = \phi_\s p_\s$ are concentrations. The self-diffusion coefficent can be identified from the pre-factor of $\nabla p_\ta$. Using that the environment is stationary ($\rho_{\s} \equiv \phi_\s$), the self-diffusion of the tagged particle for $\phi\ll 1$ is
\begin{equation}
\label{equ:Ds_green}
	D_{s,g}({\boldsymbol \phi}) = D_\ta \left[ 1- \sum_{\s \in \{r,b\}} \nolimits(1+\alpha \gamma_{\ta,\s})\phi_{\s}\right]
+ O(\phi^2)  .
\end{equation}
Physically, the self-diffusion coefficient decreases with increasing excluded volume $\phi_\s$ and increasing diffusivity ratio $\gamma_{g,\s}$. Looking at the expression for $\gamma_{g,\s}$ in \eqref{gamma_def}, this means that diffusion in a slow (or even fixed, if $D_\s \equiv 0$) environment is harder than in a fast environment.

Choosing the tagged particle to be coloured red (so that it evolves like any other particle in the red species), we define the truncation of the asymptotic series \eqref{equ:Ds_green} as
\begin{equation}
\label{equ:DsneqD}
	D^\text{low}_{s,r}(\boldsymbol{\phi}) =  D_r \left[(1- \sum_{\s\in\{r,b\}} \nolimits (1+\alpha \gamma_{r,\s})\phi_{\s}\right]  .
\end{equation}

Performing the analogous computation on the mean-field discrete PDE \eqref{limit_p_mfa}, we find that the mean-field approximation of the self-diffusion coefficient, for a tagged red particle is
\begin{equation}
\label{equ:Ds-mf}
	D_{s,r}^\text{mf}( \phi ) = D_r (1- \phi)  ,
\end{equation}
so it only depends on the total volume fraction $\phi = \phi_r + \phi_b$. 

We perform numerical simulations to test these two approximations for $D_{s,r}(\boldsymbol{\phi})$.  
 At the beginning of each simulation, we randomly populate the lattice of size $L \times L$, where $h = 1/L$, with $N_r$ red particles and $N_b$ blue particles such that the probability that a site is occupied by a red particle is $\phi_r$ and a blue particle is $\phi_b$. 
 Our stochastic model is the multi-species SSEP with jump rates \eqref{equ:lambda-lambda} and $V_\s =0$. 
 A Gillespie algorithm is used to advance the simulation until we reach some predetermined time $t = T$. Our algorithm first considers all $2Nd$ possible jumps, including those that would break the particle exclusion rule. Based on these jump rates, a random jump time is generated, and a jump is sampled. The time elapsed, $t$, is increased to the next jump time. If the chosen jump is to an empty site, the particle jump is executed, and if occupied, no jump is executed. This repeats until $t \geq T$.

\begin{figure}
\centering
	\includegraphics[width=\textwidth]{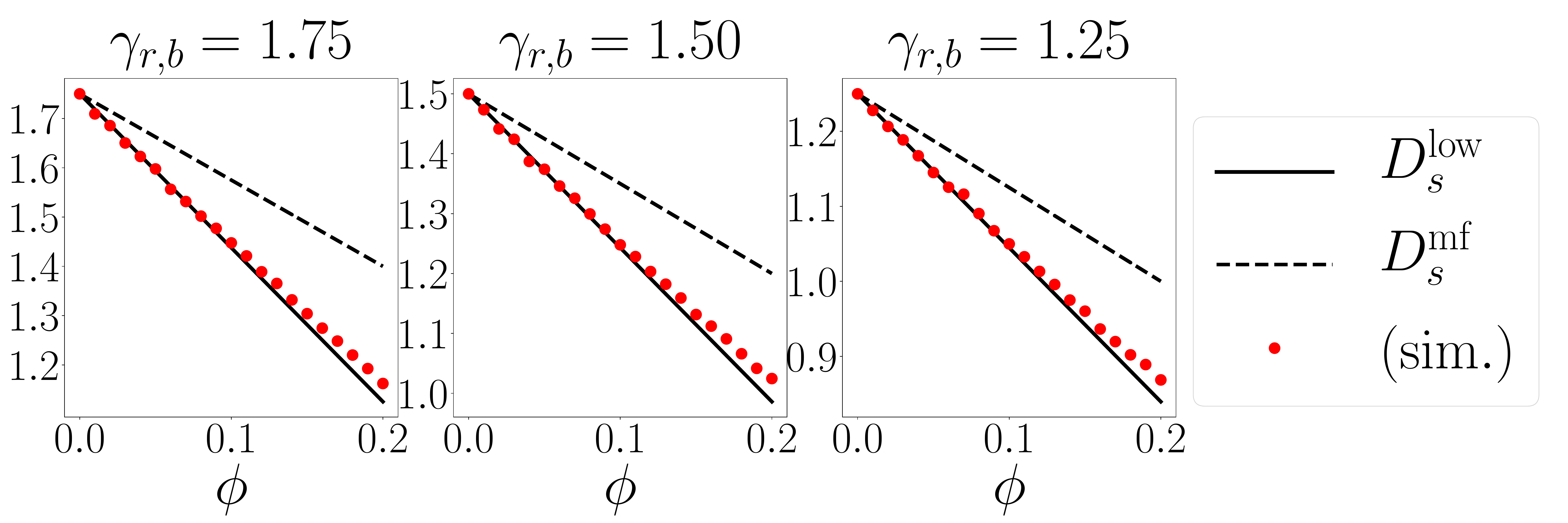}
	\caption{Dependence of the self-diffusion coefficient $D_{s,r}(\phi_r, \phi_b)$ \eqref{equ:self-diff} on the particle concentration $\phi$ and $\gamma_{r,b}$, up to the concentration $\phi = 0.2$. Results from stochastic simulation on a $10^3 \times 10^3$ periodic lattice, $h = 0.001$, $\phi_r = \phi_b$, $D_0 =1$, averaged over $K= 10$ realisations, with varying $N$ and $\gamma_{r,b}$. Measured values from stochastic simulations \eqref{equ:self-diff} (red circles) and theoretical predictions, $D^\text{low}_{s,r}$ \eqref{equ:DsneqD} (solid line) and $D^\text{mf}_{s,r}$ \eqref{equ:Ds-mf} (dashed line). }
	\label{fig:dsig vs rho}
\end{figure}

We calculate $D_{s,r}(\boldsymbol{\phi})$ numerically by averaging $\vert X_t-X_0\vert ^2/(2 d t)$ over all the red particles, over $K=10$ realisations, and over sufficiently large inspection times, $T \in [250,300]$. 
We estimate $D_{s,r}(\boldsymbol{\phi})$ numerically following this procedure for varying total occupied fraction $\phi \in [0, 0.2]$  and diffusivities ratio $\gamma_{r,b}$ (whilst keeping $D_r+D_b = 2$ and $\phi_r = \phi_b$ fixed). The results are shown as red circles in Fig. \ref{fig:dsig vs rho}. The truncation $D^\text{low}_{s,r}(\boldsymbol{\phi})$, shown as a black line, compares well with the measured values for low volume fractions and performs far better than the mean-field curve $D^{\text{mf}}_{s,r}$, shown as a black dashed-line. 

%%%%%%%%%%%%%%%%%%%%%%%%%%%%%%%%%%%%
\subsection{Self-diffusion for \texorpdfstring{$D_r =D_b$}{Dreq =
Db} }
\label{sec:self-diff:same}
%%%%%%%%%%%%%%%%%%%%%%%%%%%%%%%%%%%%

In contrast to the case  $D_r \ne D_b$, the general behaviour of the self-diffusion coefficient for the uniform environment ($D_r = D_b$) has been widely studied \cite{lamdim,Sophn}. Without loss of generality, we set $D_r=D_b=1$ and denote the self-diffusion in this case simply by $D_s(\phi)$, dropping the $\s$ index. While $D_s(\phi) \in C^\infty([0,1])$ for $\phi\in [0,1]$ \cite{lamdim}, its dependence on $\phi$ is not known explicitly, it is given instead by a variational formula \cite{Varbook}. 
However, Remark 5.3 in \cite{lamdim} provides a method for computing $D_s(\phi)$ as a Taylor expansion about either $\phi=0$ or $\phi=1$. We show in Appendix \ref{appendix:Ds} that application of this method yields the expansions \eqref{equ:Dexpan linear}, whose first-order truncations in $\phi$ and $1-\phi$ respectively are given by  
\begin{align}
    D^\text{low}_{s}(\phi) =  
    1-(1+\alpha) \phi 
      , 
	&&	
	D^\text{high}_{s}(\phi) =  
	 \frac{1}{2\alpha+1}(1-\phi)  .
	\label{equ:Dexpan_Labels}
\end{align}

In order to validate our asymptotic results, we now combine the method of \cite{lamdim} with the hydrodynamic PDE system obtained by Quastel \cite{Quastel}, given by the gradient-flow structure \eqref{equ:gen-hydro} with mobility and energy \eqref{quastel}. In particular, we substitute the low-volume approximation $D^\text{low}_{s}(\phi)$  \eqref{equ:Dexpan_Labels} into the mobility matrix \eqref{equ:M-quas}.
The result is
\begin{equation}
		\p_t \rho_\sigma =  \nabla \cdot \Big[ (1-\rho) \nabla \rho_\sigma + \rho_\sigma \nabla \rho 
		+ \alpha  \rho_{\os} \nabla \rho_\sigma  - \alpha \rho_\sigma \nabla \rho_{\os}  \Big] + O(\rho^3)   .
\end{equation}
which agrees with the matched asymptotics result \eqref{eq:mainresult} with $D_r = D_b$.  In other words, our asymptotic derivation predicts the correct behaviour of the rigorous hydrodynamic limit \cite{Quastel} up to the expected order. As discussed in  Subsection~\ref{appendix: dependence}, the agreement between the two methods can be explained by the connection between the inner problem of the matched asymptotic analysis and the recursive problems arising in the variational characterisation of $D_s$~\cite{lamdim}. 

\begin{figure}[hbt]
\centering
\includegraphics[width=0.5\textwidth]{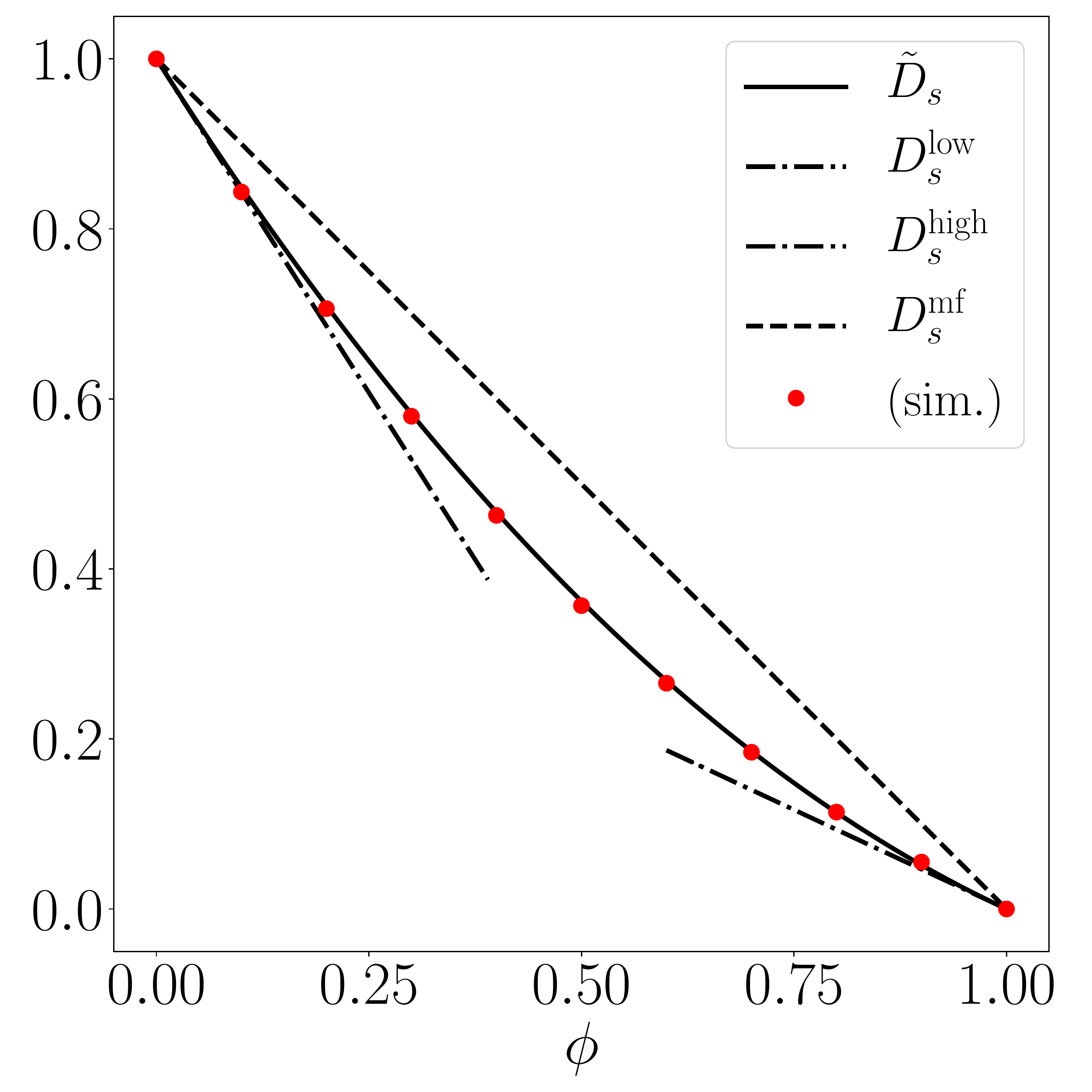}
\caption{Dependence of the self-diffusion coefficient $D_s(\phi)$ \eqref{equ:self-diff} on the occupied fraction $\phi$. Values shown are: simulated values (red circles) measured with \eqref{equ:self-diff}, composite approximation $\tw{D}_s$ \eqref{equ:Dsapprox} (solid line), mean-field approximation $D^\text{mf}_s$ \eqref{equ:Ds-mf} (dashed line), and low- and high-density approximations $D^\text{low}_{s}$ and $D^\text{high}_{s}$ \eqref{equ:Dexpan_Labels} (dot-dashed lines). Stochastic simulations performed on a $10^3 \times 10^3$ periodic lattice, for $h = 0.001$ fixed while varying $N$ such that $Nh^2 = \phi$.}
\label{fig:ds vs rho}
\end{figure}

Finally, by combining the low- and high-volume asymptotics of $D_s(\phi)$ \eqref{equ:Dexpan_Labels}, we obtain the following (minimal) cubic polynomial approximation $\tw{D}_s(\phi)$ that matches $D^\text{low}_{s}$ and $D^\text{high}_{s}$ at both ends $\phi= 0,1$ respectively:
\begin{equation}
	\label{equ:Dsapprox}
	\tw{D}_s (\phi ) = (1- \phi) \left[ 1 - \alpha  \phi + \dfrac{\alpha(2\alpha -1)}{2\alpha+1} \phi^2  \right]  .
\end{equation}
We call this the composite approximation to emphasise that it interpolates between the low and high density asymptotic expansions \eqref{equ:Dexpan_Labels}.
We numerically estimate $D_s(\phi)$ using the same procedure as in Subsection~\ref{sec:self-diff:Diff}, now with $D_r = D_b =1$ and for $\phi\in [0,1)$. 
 We find that the composite approximation $\tw{D}_s$ agrees extremely well with the simulated data (see Fig.~\ref{fig:ds vs rho}). This contrasts with the mean-field approximation $D^{\text{mf}}_{s}$.

%%%%%%%%%%%%%%%%%%%%%%%%%%%%%%%%%%%%
\section{Numerical simulations of collective dynamics}  \label{sec:numerics}
%%%%%%%%%%%%%%%%%%%%%%%%%%%%%%%%%%%%

In this section, we perform numerical simulations of the cross-diffusion PDE system \eqref{equ:gen-hydro} for the concentrations $\rho_r$ and $\rho_b$,  with the mobility matrix $M$ obtained either via matched asymptotics \eqref{equ:M-mix}, mean-field \eqref{equ:M-mf} or the composite approximation of \eqref{equ:M-quas} obtained via \eqref{equ:Dsapprox}. The microscopic model (simple exclusion process) described in Section \ref{sec:model} is used to benchmark and test the validity of such PDE models. 

As in the previous section, the microscopic model is simulated using the Gillespie algorithm \cite{stochbio}, while the PDE models are solved using a discretisation of the gradient flow \eqref{equ:gen-hydro}. The scheme is chosen so that in the cases that a discretised equation already exists (\eqref{integrated_eq} and \eqref{equ:mf-disc}), the equations will match to $O(h,\phi,\Delta t)$. For a small time step $\Delta t \sim h^2$, we consider the numerical scheme
\begin{equation} \label{equ:gen scheme}
\renewcommand*{\arraystretch}{1.1}
\begin{pmatrix}
\rho_r(x, t+\Delta t ) \\
\rho_b(x, t+\Delta t )
\end{pmatrix} = \begin{pmatrix}
\rho_r(x, t) \\
\rho_b(x, t )
\end{pmatrix}+ \frac{\Delta t}{h^2} \sum_{k} [J_{x+he_k/2}(t)-J_{x-he_k/2}(t)],
\end{equation}
with periodic boundary conditions and
\begin{equation}
\renewcommand*{\arraystretch}{1.3}
    J_{x+he_k/2} =  M(\rho_r, \rho_b)     \begin{pmatrix}  {\bar F}_r(x+\tfrac{h}{2}e_k) \\ {\bar F}_b(x + \tfrac{h}{2}e_k) \end{pmatrix},
\end{equation}
where the discretised thermodynamic force is,
\begin{equation} \label{equ:Ediscrete}
    {\bar F}_\s(x+\tfrac{h}{2}e_k) = \frac{\rho_\s(x+h e_k)-\rho_\s(x)}{h\rho_\s(x+\tfrac{h}{2}e_k)}+ \frac{\rho(x+he_k)-\rho(x)}{h[1- \rho(x+\tfrac{h}{2}e_k)]} + \nabla V_\s(x+\tfrac{h}{2}e_k),
\end{equation}
where $\rho(x+\tfrac{h}{2}e_k) = [\rho(x+he_k) + \rho(x)]/2$.

We present simulations in two dimensions, with $\Omega = \{ (h\ZZ)^2 \cap [0,1)^2 \}$ and periodic boundary conditions.
Initially red particles are distributed uniformly on $\{ x \in \Omega: x_1 \leq 1/2\}$, $\rho_r(\cdot,0) = \phi \cdot \id_{(0,1/2]}$ and blue particles on $\{ x \in \Omega: x_1 > 1/2\}$, $\rho_b(\cdot,0) = \phi \cdot \id_{(1/2,1]}$. 
We set the potentials to $ D_r V_r = -D_b V_b = \sin (2 \pi x) $, and have a lattice spacing  $h =0.01$.
Due to the vertical symmetry of the system $\rho_r$ and $\rho_b$, are constant in the $x_2$ direction. At a given time $t$, we construct the average density profiles, $\langle \eta_t^r(x) \rangle $ and $\langle \eta_t^b(x) \rangle $, by averaging over the $x_2$ coordinates and over $K$ realisations. We plot histograms of $\langle \eta_t^r(x) \rangle$, $\langle \eta_t^b(x) \rangle$ with bin width $w$ (red/blue circles). 

%%%%%%%%%%%%%%%%%%%%%%%%%%%%%%%%%%%%
\subsection{Case \texorpdfstring{$D_r = D_b$}{Dr = Db} }
%%%%%%%%%%%%%%%%%%%%%%%%%%%%%%%%%%%%

In the case $D_r = D_b$ we benchmark two PDE models against the numerical simulations. In the case $D_r = D_b$ we can still use the mobility $M^\text{sym}$ from the rigorous hydrodynamic limit.
It is noted in \cite{Erignoux} that the nongradient method of \cite{Quastel,Erignoux} is compatible with a smooth potential. Hence we obtain a hydrodynamic equation by replacing \eqref{equ:E0} with the general hydrodynamic force \eqref{equ:Ediscrete}.
However, the density-dependent self-diffusion coefficient $D_s$ that appears in $M^\text{sym}$ \eqref{equ:M-quas} is not known explicitly, so it must be approximated to obtain an equation that can be solved numerically.  
This is achieved by 
replacing $D_s$ in the mobility $M^\text{sym}$ by the composite approximation $\tw{D}_s$ \eqref{equ:Dsapprox}. 
We compare the performance of \eqref{equ:gen scheme} with the mobility $M^\text{sym}$ and approximate diffusion coefficient with solutions denoted, ($\tw{\rho}_r, \tw{\rho}_b$), to the solutions with the mean field mobility $M^\text{mf}$ denoted, ($\rho_r^\text{mf}, \rho_b^\text{mf} $).  

\begin{figure}[t]
\centering
	\includegraphics[width=\textwidth]{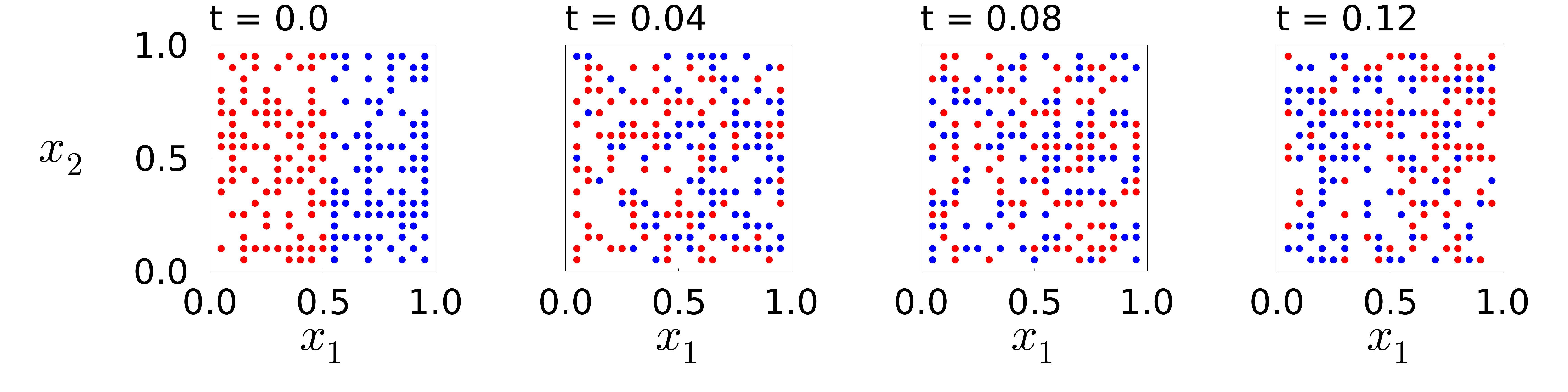}
	\caption{Visualisation of particle positions, for $D_r = D_b =1$, $V_r = -V_b = \sin (2 \pi x) $,  $h =0.05$ and $\phi_r = \phi_b = 0.25$ ($N_r = N_b = 100$). Snapshots of the process are taken at $t= 0.0,~0.04, ~0.08,~ 0.12$}
	\label{fig:snapshot}
\end{figure}

\begin{figure}[bt!]
\centering
	\includegraphics[width=.9\textwidth]{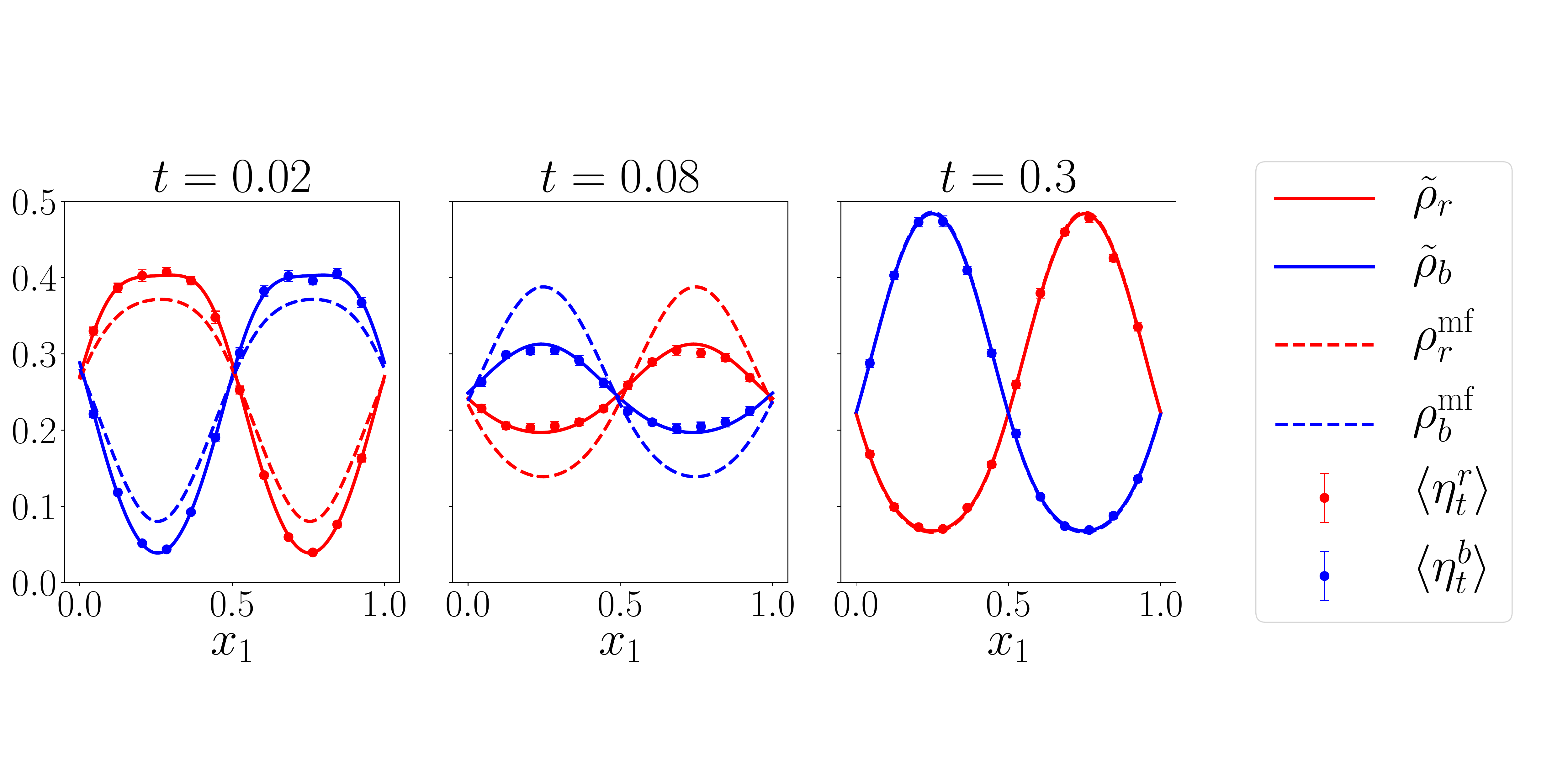}
	\caption{
		Particle densities at times \textcolor{black}{$t = 0.02, 0.08, 0.3$} with initial data $\rho_r(\cdot,0) = \frac{1}{2} \id_{(0,1/2]}$, $\rho_b(\cdot,0) = \frac{1}{2}\id_{(1/2,1]}$ and parameters $D_r = D_b =1$, $V_r = -V_b = \sin (2 \pi x) $,  $h =0.01$ and $\phi_r = \phi_b = 0.25$ ($N_r = N_b = 2500$). Solutions of \eqref{equ:gen-hydro}, $\tw{\rho}_r, \tw{\rho}_b$ (solid line) and  $\rho_r^\text{mf}, \rho_b^\text{mf}$ (dashed line), and histograms of $\langle \eta_t^r(x) \rangle$, $\langle \eta_t^b(x) \rangle$ (data points) averaged over $x_2$ and $K = 30$ realisations, where the density of each species is shown in its respective colour. Error bars indicate twice the standard error.
	}
	\label{fig:rho vs x}
\end{figure}

To test $\tw{D}_s$ furthest from its asymptotic derivation, we set $\phi = 0.5$. We have performed simulations of the two-species model with $D_r = D_b =1$. Snapshots of a reduced system ($h= 0.05$) are shown in Fig. \ref{fig:snapshot}. 
In Figure \ref{fig:rho vs x} we plot histograms of $\langle \eta_t^r(x) \rangle$, $\langle \eta_t^b(x) \rangle$ from $K=30$ realisations, with bin width $w = 0.08$,  (red/blue circles).
 

The averaged discrete densities, $\langle \eta_t^r(x) \rangle$, $\langle \eta_t^b(x) \rangle$  compare well with the approximate solutions $\tw{\rho}_r, \tw{\rho}_b$ using the composite approximation $\tw{D}_s ( \phi ) $ for the self-diffusion coefficient (Fig. \ref{fig:rho vs x}).
The diffusive nature of the particles causes the initial blocks of red and blue particles to spread out over time, and furthermore the potential $V_r = \sin (2 \pi x) $ ($V_b = - \sin (2 \pi x) $) pushes the red (blue) particles over to the right (left). These factors act together to transport particles and later balance in the steady state.
The mean-field equation \eqref{equ:2body} has a higher mobility  than the composite prediction ($\tw{D}_s ( \phi ) < D^\text{mf}_s ( \phi )$) so it appears that the mean-field solution ($\rho_r^\text{mf}, \rho_b^\text{mf} $) is significantly ahead of $\tw{\rho}_r, \tw{\rho}_b$ and $\langle \eta_t^r(x) \rangle$, $\langle \eta_t^b(x) \rangle$ ($t= 0.02$ and $t= 0.08$).
At later times however both solutions, $\rho_r^\text{mf}, \rho_b^\text{mf} $ and $\tw{\rho}_r, \tw{\rho}_b$ converge to the same steady state ($t = 0.3$) because they have the same free energy \eqref{equ:Etot}. 

\begin{figure}[tb]
\centering
	\includegraphics[width=0.65\textwidth]{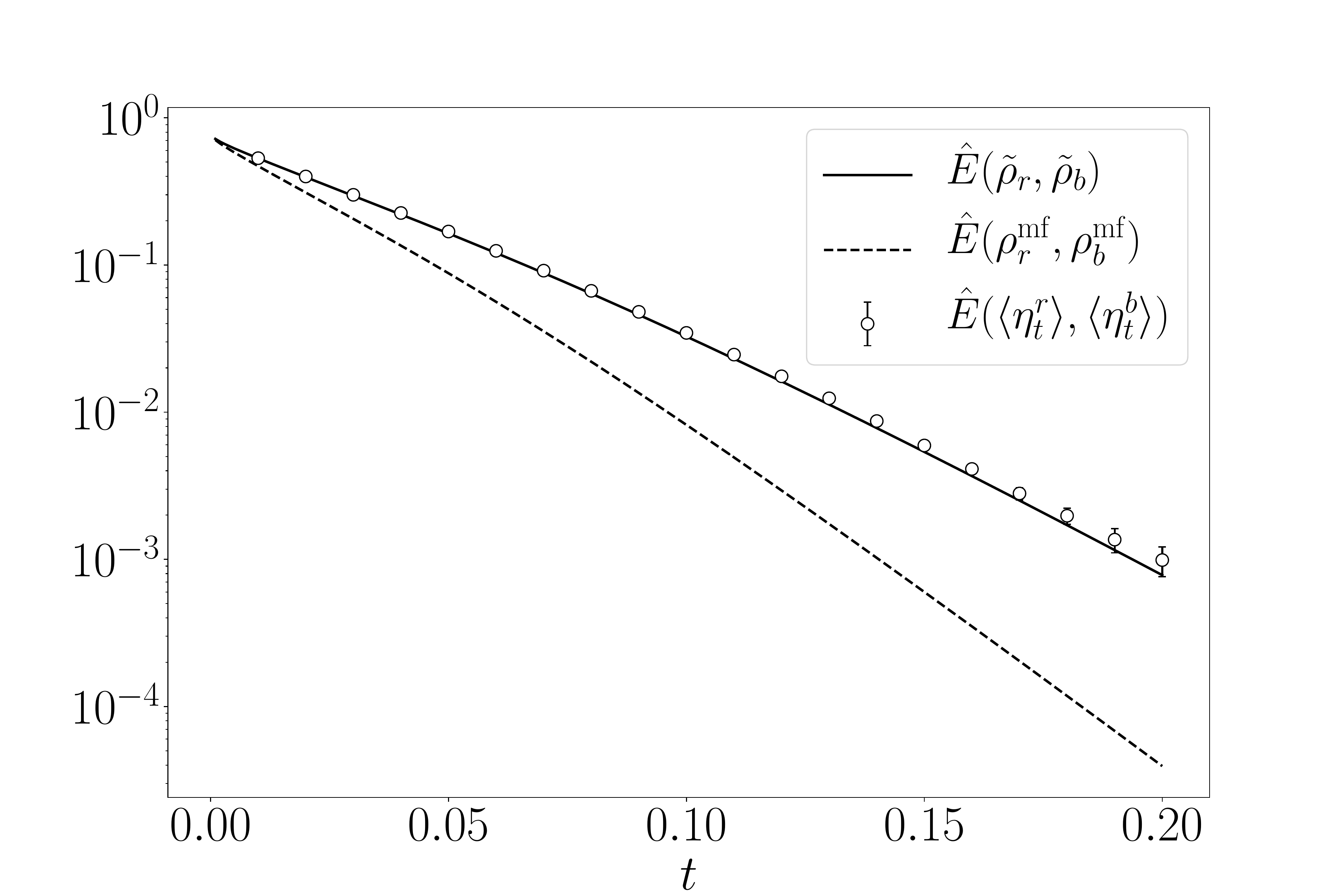}
		\caption{
		Relative free energy \eqref{equ:Etot} for $0.00 < t < 0.20$ and parameters $D_r = D_b =1$, $V_r = -V_b = \sin (2 \pi x) $,  $h =0.01$ and $\phi_r = \phi_b = 0.25$ ($N_r = N_b = 2500$). Comparison between the free energy evaluated with solutions of \eqref{equ:gen-hydro},  $\tw{\rho}_r, \tw{\rho}_b$ (solid line) and  $\rho_r^\text{mf}, \rho_b^\text{mf}$ (dashed line), or via simulated data, $\langle \eta_t^r(x) \rangle$, $\langle \eta_t^b(x) \rangle$, averaged over $x_2$ and $K = 60$ realisations (data points). Error bars indicate twice the standard error, and are initially covered by the data points.}
	\label{fig:E vs t}
\end{figure} 
The free energy functional $E$ \eqref{equ:Etot} has a unique minimiser $(\rho_r^\infty,\rho_b^\infty)$ \cite{Burger}, which corresponds to the steady state of the cross-diffusion system (note that all PDE models we discuss share the same steady state as they have the same free energy). We consider the relative free energy
\beq
\hat{E}(t) = E[\rho_r, \rho_b](t) - E[\rho_r, \rho_b](\infty),
\label{equ:hat-E}
\eeq 
in Figure \ref{fig:E vs t}, with $(\rho_r,\rho_b)$ taken from either the composite approximation model, mean field or stochastic simulations.

Furthermore, the free energy \eqref{equ:Etot} can be calculated using the density profiles. As the concentration of opposite particles decreases, the mobility of a particle increases. Hence in Fig. \ref{fig:E vs t}, instead of a straight line, we see a curve with varying gradient. 
In Fig. \ref{fig:rho vs x} we can see that initially, particles are required to move through a high density of the opposing species to reach the steady-state as the red (blue) must travel from left (right) to the right (left). 
However, transport through the opposing species is reduced for later times when the system is closer to equilibrium. This leads to a steeper slope in Fig. \ref{fig:E vs t} at longer times. 
 
We remark that the estimation of the time-dependent free energy \eqref{equ:hat-E} from $\tilde{\rho}_\s$ is biased, because 
fluctuations in $\tilde{\rho}_\s=\langle \eta_t^\s(x) \rangle$, lead to a systematic increase in $\hat{E}$ at $O( \text{Var} \langle\eta_t^\s(x)\rangle )$.
However, by ensuring the variance of the free energy is low, we see excellent agreement between the free energy of the asymptotic solution and the free energy evaluated using the average density profiles. On the other hand, the mean-field solution has greater mobility and moves faster towards equilibrium.  

%%%%%%%%%%%%%%%%%%%%%%%%%%%%%%%%%%%%
\subsection{Collective dynamics: \texorpdfstring{$D_r \neq D_b$}{DrneqDb} }
%%%%%%%%%%%%%%%%%%%%%%%%%%%%%%%%%%%%

In the general setting ($D_r \neq D_b$), we lack a rigorous proof of the probabilistic convergence of paths as $h \rightarrow 0$  and furthermore, lack not only a a global approximation of $D_{s,\s}( \boldsymbol{\phi})$  but also rigorous proof of its existence and regularity. Therefore to model particle density we turn to the asymptotic PDE \eqref{equ:asymp soln} and use stochastic simulations as a benchmark. We denote the solutions to \eqref{equ:gen scheme} with asymptotic mobility $M^\text{low}$ by $\rho_r^\text{low}, \rho_b^\text{low}$ and solutions with the mean field mobility $M^\text{mf}$ by $\rho_r^\text{mf}, \rho_b^\text{mf}$.  

\begin{figure}[bt!]
\centering
	\includegraphics[width=.9\textwidth]{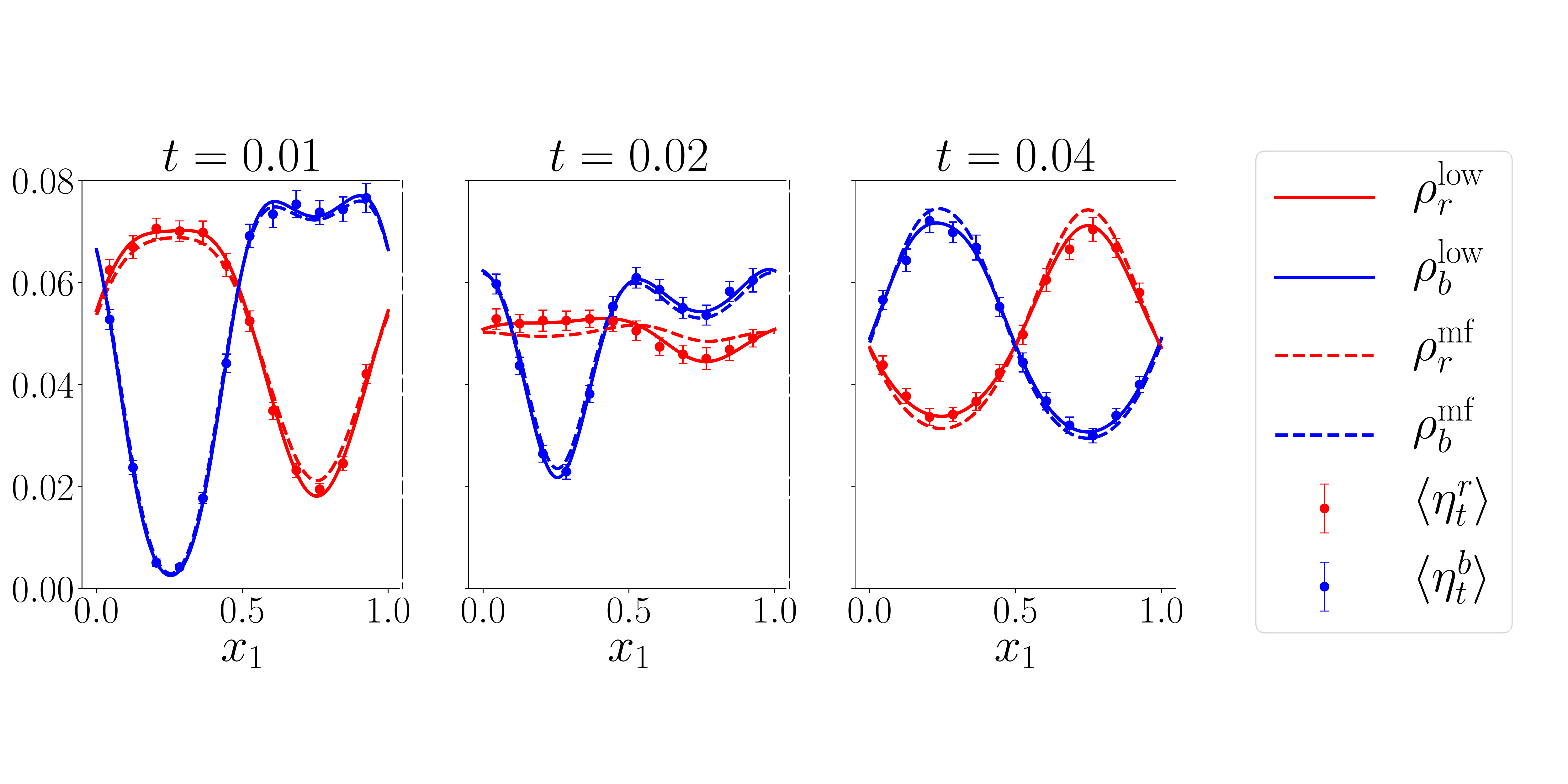}
	\caption{
			Particle densities at times \textcolor{black}{$t = 0.01, 0.02, 0.04$} with initial data $\rho_r(\cdot,0) = 0.1 \id_{(0,1/2]}$, $\rho_b(\cdot,0) = 0.1\id_{(1/2,1]}$ and parameters \textcolor{black}{$D_r = 1.5$, $D_b=0.5$}, $ D_r V_r = - D_b V_b = \sin (2 \pi x) $,  $h =0.01$ and $\phi_r = \phi_b = 0.05$ ($N_r = N_b = 500$). Solution of \eqref{equ:asymp soln}  $\rho_r^\text{low}, \rho_b^\text{low}$(solid line), solution of \eqref{equ:2body} $\rho_r^\text{mf}, \rho_b^\text{mf}$ (dashed line) and histograms of $\langle \eta_t^r(x) \rangle$, $\langle \eta_t^b(x) \rangle$ averaged over $x_2$ and $K = 60$ realisations (data points), where the density of each species is shown in its respective colour. Error bars indicate twice the standard error. }
\label{fig:diff rho vs x}
\end{figure}

Figure \ref{fig:diff rho vs x} shows the results of simulations with $D_r = 1.5$, $D_r = 0.5$ and $\phi_r = \phi_b = 0.05$. We plot histograms of $\langle \eta_t^r(x) \rangle$, $\langle \eta_t^b(x) \rangle$ from $K=60$ realisations, with bin width $w = 0.08$ (red/blue circles). The averaged discrete densities, $\langle \eta_t^r(x) \rangle$, $\langle \eta_t^b(x) \rangle$  compare well with the asymptotic solutions $\rho_r^\text{low}, \rho_b^\text{low}$ (Fig. \ref{fig:diff rho vs x}).

Initially, red particles are placed uniformly on the left half of the periodic lattice and blue on the right.
The diffusive nature of the particles causes the block of red (blue) particles to spread out over time, and furthermore the potential $D_r V_r = \sin (2 \pi x) $ ($D_b V_b = - \sin (2 \pi x) $) pushes the red (blue) particles over to the right (left). Initially these factors act together to transport particles and later balance out in the steady state.
In the blue particles, we see an additional local minimum form ($t=0.1,0.2$) as the potential is strong enough compared to the diffusion constant. In contrast, this bump is spread out by stronger diffusion in the red particles.
The mean field equation \eqref{equ:2body} has a higher mobility  than the asymptotic equation \eqref{equ:asymp soln} so it appears that the mean-field solution $\rho_r^\text{mf}, \rho_b^\text{mf} $ is  ahead of $\rho_r^\text{low}, \rho_b^\text{low}$ and $\langle \eta_t^r(x) \rangle$, $\langle \eta_t^b(x) \rangle$.
This difference only appears at intermediate times however as both solutions, $\rho_r^\text{mf}, \rho_b^\text{mf} $ and $\rho_r^\text{low}, \rho_b^\text{low}$ converge to the same steady state because they have the same free energy \eqref{equ:Etot}.

%%%%%%%%%%%%%%%%%%%%%%%%%%%%%%%%%%%%
\section{Discussion} \label{sec: discussion}
%%%%%%%%%%%%%%%%%%%%%%%%%%%%%%%%%%%%

We have considered a two-species SEP on a periodic lattice: this is a discrete model for the diffusion of particle mixtures in the presence of an external potential.
Our analysis uses the method of matched asymptotics, which yields a set of coupled ODEs for the (lattice-based) probability distribution density of the particle positions $P_\sigma(x,t)$ for each species. We interpret this as a discrete cross-diffusion PDE system, from which we identify corresponding macroscopic PDEs.
These can then be factorised into the gradient-flow structure \eqref{equ:gen-hydro}. Within this description, a non-trivial mobility matrix $M$ captures the effect of microscopic interactions on the collective dynamics, which manifest as coupled nonlinear cross-diffusion terms.

We compare the predictions of this PDE system with numerical simulations of the discrete stochastic model. Specifically, we consider self-diffusion and the collective dynamics of the particle densities. For small volume fractions $\phi$, the predictions of the PDE show excellent agreement with the microscopic model, confirming that it captures the macroscopic behaviour accurately. This quantitative agreement can be attributed to the systematic nature of the method of matched asymptotics, which relies on an expansion for small lattice parameter $h$ and small occupied volume fraction $\phi$. As $\phi\to0$, particle exclusion interactions are negligible; the expansion over $\phi$ enables these interactions to be characterised, order by order.  

We have focused on the first non-trivial order in this expansion, which yields \eqref{eq:mainresult}. (The generalisation to higher order is discussed in Appendix~\ref{appendix:matched second}.) The dominant effect of interactions -- illustrated by the numerical results of Sec.~\ref{sec:numerics} -- is that the presence of the blue species slows down the collective diffusion of the red species (and vice versa). Physically, this is natural because the blue particles act to block the motion of red ones via the exclusion constraint. This effect is underestimated by theories that rely on mean-field closure of the equations of motion~\cite{Burger,mean1}.

Notably, the results obtained for the SEP lattice-based model differ significantly from off-lattice models~\cite{Maria2}, where the collective diffusion is enhanced by interactions between particles of the same species. To understand this, note that the invariant measure of the off-lattice system incorporates non-trivial correlations between particles, due to their packing in space. Such effects enter the free energy through the virial expansion~\cite{BBGKY}. These packing constraints become increasingly important at high volume fractions, increasing the pressure and accelerating diffusion away from high-density regions. There are no such packing effects in the SEP: the invariant measure has a product structure and the free energy \eqref{equ:Etot} is exact for all $\phi$. In this case, the interaction terms from the matched asymptotic expansion only appear in the mobility matrix $M$, which act to suppress diffusive spreading. 

Compared with previous results for hydrodynamic limits, our model extends previous work~\cite{Quastel,Erignoux,Varbook} by allowing the two-particle species to have different diffusion constants, $D_r\neq D_b$ (as well as different drifts). For the particular case $D_r=D_b$, our macroscopic PDEs match the ones derived rigorously in~\cite{Quastel,Erignoux}, up to the expected order in $\phi$. We also showed (for $D_r=D_b$) that the matched asymptotics approach can be used to derive the self-diffusion constant that agrees with the expansion in $\phi$ that was derived rigorously in~\cite{lamdim}. To the best of our knowledge, our results for multi-species SEP with $D_r\neq D_b$ are the first to give a consistent picture of the collective dynamics and self-diffusion.

Rigorous analysis of the hydrodynamic limit is technically challenging in these multi-species systems because they are of non-gradient type~\cite{Quastel,Erignoux,Varbook}. In particular, one requires a sharp estimate of the spectral gap to establish the system's fast-mixing properties.  
We note that recent results for the spectral gap are available for multi-species exclusion processes~\cite{boundpaper}, which might enable a rigorous analysis of the hydrodynamic limit for $D_r\neq D_s$.

In this context, the role of the matched asymptotic method is to provide a simple and physically intuitive route to the hydrodynamic PDEs: the method avoids the technical difficulties in proving fast mixing, and hence convergence in probability of the particle densities. It also allows lattice-based and off-lattice systems~\cite{Maria2,Maria1,vJump} to be considered on an equal footing. Given this versatility, it will be interesting to explore future applications to other lattice systems of non-gradient type, including those for which spectral gap estimates are not available. This might include lattice models that incorporate packing effects, in order to make contact with off-lattice results such as~\cite{Maria2}.

%%%%%%%%%%%%%%%%%
\subsection*{Acknowledgements}
%%%%%%%%%%%%%%%%%

M.~Bruna was supported by a Royal Society University Research Fellowship (grant no.~URF/R1/180040). J.~Mason was supported by the Royal Society Award (RGF/EA/181043) and the Cantab Capital Institute for the Mathematics of Information of the University of Cambridge. The authors thank Tal Agranov, Mike Cates and Jon Chapman for helpful discussions.

%%%%%%%%%%%%%%%%%
\subsection*{Data availability}
%%%%%%%%%%%%%%%%%
 
The datasets generated during the current study are available from the corresponding author on reasonable request. 

\begin{appendices}

%%%%%%%%%%%%%%%%%%%%%%%%%%%%%%%%%%%%
\section{Outer Solution}
\label{app:Outer Solution}
%%%%%%%%%%%%%%%%%%%%%%%%%%%%%%%%%%%%

In this appendix we prove the relationship \eqref{equ:outer lead}. Recall we established $P_{\text{out}}(x,y,t) = Q_\sigma(x,t) Q_{\sigma'}(y,t)+O(\phi)$ for some functions $Q_\sigma$ satisfying $\dot Q_\s = \mathcal{L}^*_\s Q_\sigma$. 
%And therefore it follows from 
Hence \eqref{inner0} implies $P_{\text{in}}(\u,\v,t) = Q_\sigma(\u,t) Q_{\sigma'}(\u,t)+O(h,\phi)$. 
 
Although there is no definite boundary between the inner and outer region, here it is useful to fix some radius, $\ep$, so that $h \ll \ep \ll 1$ and
\newcommand{\out}{\Omega^x_\text{out}}
\newcommand{\inner}{ \Omega^x_\text{in}}
\begin{align}
\begin{aligned}
	 \out   &= \{ y \in \Omega : \vert x-y \vert > \ep \} \\ 
	 \inner &= \{ y \in \Omega : 0< \vert x-y \vert \leq \ep \} .
\end{aligned}
\end{align}
By definition, the two particle density must sum to $h^{-2d}$ over its domain. Splitting the sum over $y$ into the inner and outer region yields
\begin{equation}
		1  
	= \sum_{x,y \in \Omega} h^{2d} P_{\s, \s'} 
	= 
	\sum_{\substack { x \in \Omega \\
	y \in \out } } 
	h^{2d} P_\text{out}
	+ 
	\sum_{\substack { x \in \Omega \\
	y \in \inner } }
	h^{2d} P_\text{in} ,
\end{equation}
where it its implict that $P_\text{in}$ is evaluated using the respective inner variables.
Now we substitute expressions for the inner and outer solutions, 
\begin{align*}
	1
	&=
	\sum_{x \in \Omega}h^{d} Q_{\s} \Bigg(\sum_{y \in \out}h^{d} Q_{\s'}(y) + \sum_{y \in \inner} h^{d} Q_{\s'}(x) + O(h \ep^{d}) \Bigg)
		\\
	&=
	\sum_{x \in \Omega}h^{d} Q_{\s} 
	\Bigg(\sum_{y \in \Omega}h^{d} Q_{\s'}(y) + \sum_{y \in \inner} h^{d} \Big( Q_{\s'}(x) - Q_{\s'}(y) \Big) + O(h \ep^{d}) \Bigg) \nonumber
	\\
	&=
	\sum_{x \in \Omega}h^{d} Q_{\s} \sum_{y \in \Omega}h^{d} Q_{\s'}(y) + O(h \ep^{d}), 
\end{align*}
where we use that $\sum_{y \in \inner}h^d$ is $O(\ep^d)$. Therefore without loss of generality we can take $\sum_{x \in \Omega}h^{d} Q_{\s}  = 1 + O(h \ep^{d})$. 

By definition, summing over $y \in \Omega$ the two particle density must sum to $h^{-d} P_{\s}(x,t)$. Splitting the sum over $y$ into the inner and outer region, one obtains
\begin{equation}
		P_{\s} (x,t) 
	= \sum_{y \in \Omega} h^{d} P_{\s, \s'} (x,y,t)
	= 
	\sum_{
	y \in \out } 
	h^{d} P_\text{out} (x,y,t)
	+ 
	\sum_{
	y \in \inner } 
	h^{d} P_\text{in} (x,\v,t) .
\end{equation}
Now we substitute expressions for the inner and outer solutions, 
\begin{align*}
	P_{\s} (x,t)
	&=
	Q_{\s} (x,t)\Bigg(\sum_{y \in \out}h^{d} Q_{\s'}(y) + \sum_{y \in \inner} h^{d} Q_{\s'}(x) + O(h \ep^{d}) \Bigg)
		\\
	&=
	 Q_{\s}(x,t) \Bigg(\sum_{y \in \Omega}h^{d} Q_{\s'}(y) + \sum_{y \in \inner} h^{d} \Big( Q_{\s'}(x) - Q_{\s'}(y) \Big) + O(h \ep^{d}) \Bigg) \nonumber
	\\
	&=
	Q_{\s}(x,t) + O(h \ep^{d}).
\end{align*}
Finally fixing $\ep = h^\frac{d-1}{d}$, it follows that $P_{\text{out}}(x,y,t) = P_\sigma(x,t) P_{\sigma'}(y,t)+O(\phi)$ as required. 
%%%%%%%%%%%%%%%%%%%%%%%%%%%%%%%%%%%%
\section{Properties of the auxiliary function \texorpdfstring{$\bgfpsi$}{psi}}
\label{app:psi}
%%%%%%%%%%%%%%%%%%%%%%%%%%%%%%%%%%%%

The $d$-vectorial function $\bgfpsi$ has components $\psi_j$ satisfying \eqref{aux_problem},  
which we recall here for convenience
\begin{subequations} \label{aux_problem2}
	\begin{alignat}{3} \label{aux_eqn}
		 \lap \psi_j  &=   \v_j   \ind ,& \quad & \v \in \ZZ^d,\\
	\psi_j &\sim 0, & &\vert \v\vert \to \infty. 
	\end{alignat}
\end{subequations}
This appendix derives some of its properties. To this end, we consider its semidiscrete Fourier transform $\bgfpsitw$, defined for  $\zeta\in [-\pi,\pi]^d$ as
\beq
\bgfpsitw(\zeta) = \sum_{\v\in \mathbb{Z}^d} \bgfpsi(\v) {\rm e}^{-i\zeta \cdot \v}.
\eeq
We recall that the semidiscrete Fourier transform takes an unbounded and discrete spatial domain ($\v \in \ZZ^d$ in our case) to a bounded and continuous frequency domain ($\zeta\in [-\pi,\pi]^d$). Taking the Fourier transform of \eqref{aux_eqn} yields
\beq
-4\gfpsitw_j(\zeta) \sum_{k=1}^d \sin^2 \frac{\zeta_k}{2} = e^{-i\zeta_j} -e^{i\zeta_j},
\eeq
where $\gfpsitw_j$ is the $j$th component of $\bgfpsitw$. Then $\psi_j$ can be recovered from $\tilde \psi_j$ via the inverse semidiscrete Fourier transform
\beq
\bgfpsi(\v) = \frac{1}{(2\pi)^d} \int_{[-\pi,\pi]^d} \bgfpsitw(\zeta)  e^{i \zeta \cdot \v } d\zeta,
\eeq
yielding
\beq
\gfpsi_j(\v) = \frac{1}{(2\pi)^d} \int_{[-\pi,\pi]^d} \frac{e^{i\zeta_j}-e^{-i\zeta_j}}{  4\sum_k \sin^2 (\zeta_k/2)} e^{i \zeta \cdot \v } d\zeta.
\label{equ:psi-integral}
\eeq

We note the following properties: 
\begin{itemize}
	\item For $\v=0$,  the integrand in \eqref{equ:psi-integral} is odd in $\zeta_j$ and so $\psi_j(\v)=0$.
	\item For $\v=e_k$ (a member of the standard basis for $\mathbb{R}^d$) the the imaginary part of the integrand in \eqref{equ:psi-integral} is odd in $\zeta_1$, which implies
\begin{align}
\gfpsi_j(e_k)  
= \beta \delta_{jk},
\end{align}
where $\beta$ was defined in \eqref{beta_def}.
\item For $\v=- e_k$, we have $\gfpsi_j(-e_k) = -\gfpsi_j(e_k) =  -\beta \delta_{jk}$.
\end{itemize}

%%%%%%%%%%%%%%%%%%%%%%%%%%%%%%%
\section{Self-diffusion coefficient of SSEP}
\label{appendix:Ds}
%%%%%%%%%%%%%%%%%%%%%%%%%%%%%%%%%%

In this appendix, we review some of the results of~\cite{lamdim} and apply them to compute the behaviour of the self-diffusion coefficient $D_s(\phi)$ as $\phi\to0$ and $\phi\to1$ in the symmetric and coloured case. In particular, \cite{lamdim} considers the evolution of a single tagged particle in the symmetric simple exclusion process (SSEP) in $\ZZ^d$ that is in equilibrium at density $\phi$. They prove that $D_s (\phi)\in C^\infty([0,1])$ and provide a recursive method of compute its Taylor expansion at the boundaries $\phi = 0,1$. It is this method that is of most relevance to us here.

%%%%%%%%%%%%%%%%%%%%%%%%%%%%%%%%%%%%
\subsection{Setup}
%%%%%%%%%%%%%%%%%%%%%%%%%%%%%%%%%%%%

The microscopic problem studied in \cite{lamdim} is defined similarly to Sec.~\ref{sec:model} but on an infinite lattice $\ZZ^d$ with unit spacing. (It is related to the process in the inner region presented in Subsec.~\ref{sec:Asymp}, as we will see below.)
 In particular, it corresponds to a two-species SSEP ($\lambda_r = \lambda_b = \id_{\vert x -y \vert}$)  with a (say) red tagged particle ($N_r = 1$) at $X_t \in \ZZ^d$ and an enviroment of blue particles (at a density $\phi_b \approx \phi$). 
 
The red particle is initialised at the origin, $X_0 = 0$. The other sites ($\zdstar = \ZZ^d \setminus \{ 0\}$) are initialised with blue particles (with probability $\phi$) or vacancies (with probability $1-\phi$). The self-diffusion of the red particle \eqref{equ:self-diff} coincides with the self-diffusion coefficient $D_s(\phi)$ in the single-species SSEP at density $\phi$ [$D_{s,r}(\boldsymbol{\phi}) \equiv D_s(\phi)$, where $\boldsymbol{\phi} = (0,\phi)$]. Note that $D_s$ can be more generally defined by the tensor \cite{lamdim,Sophn}, 
\beq
\hat{D}^{kl}_{s}(\phi) =  \lim_{t \rightarrow \infty} \frac{1}{2t} \EE \left[ ((X_t - X_0)\cdot e_k) ((X_t - X_0)\cdot e_l)  \right] .
\eeq
%note to self: to convert from lamdim paper to mine set their $p(v) = \id_{\vert v \vert =1}$ and half their $D_s$ (my $D_s$ is chosen with the convention of quastel/ Burger and looks nicer in a PDE).
from which it is clear, by the symmetry of the process, that $\hat{D}^{kl}_{s}(\phi) = D_s(\phi) \delta_{kl}$.

It is convenient to define the configuration of the environmental (blue) particles $\eta^b_t$ in the reference frame of the tagged particle \cite{lamdim}. 
The resulting environmental configuration is denoted $\xbi_t \in \{0,1\}^\zdstar$, and its corresponding site variables are related to those of the original SEP by $\xbi_t(y) = \eta^b_t(y-X_t)$ for $y \in \zdstar$.
The environmental configuration $\xbi_t$ evolves by a Markov jump process with generator $\mathcal{L}^{\rm t} = \mathcal{L}_0 +\mathcal{L}_\tau$, where the action of $\mathcal{L}_0,\mathcal{L}_\tau$ on a generic cylinder function $f$ is
\begin{align*}
	\mathcal{L}_0 f(\xbi)    &= \sum_{\substack{x,y\in \ZZ^d_*\\ \vert x-y\vert =1}}  \xi(x)[1-\xi(y)]  [ f(\xbi^{x  y}) - f(\xbi) ] = \frac{1}{2} \sum_{\substack{x,y\in \ZZ^d_*\\ \vert x-y\vert =1}}  [ f(\xbi^{x  y}) - f(\xbi) ],	\\
	 \mathcal{L}_\tau f(\xbi) &= \sum_{e \in B_0}  [1- \xbi(e)](f( \tau_{e} \xbi) -f (\xbi))	 ,
\end{align*}
%note to self: this is as if $p= 1$ in lamdim
where $\mathcal L_0$ corresponds to jumps of blue particles, while $\mathcal L_\tau$ describes the effect of a jump of the red (tagged) particle to the enviroment. Here $B_0 = \{ \v\in \zdstar : \vert \v\vert =1 \}$ and $\tau_{e} \xbi$ stands for the configuration after the tagged particle jumps by $e$ and the enviroment is shifted by $-e$:
\begin{equation}
	\tau_e \xbi (y) = \begin{cases}
 							\xbi (y + e) & \text{if } y \neq -e, \\
 							\xbi (e)  & \text{if } y = -e,
 					\end{cases} \nonumber
\end{equation}
(Note that $\xbi(e)=0$ before the jump as, otherwise, the tagged particle would have violated the size-exclusion rule.)
We also introduce $\mu_\phi$ as a Bernoulli product measure on $\{0,1\}^\zdstar$, such that $\xbi(x)=1$ with probability $\phi$ and zero otherwise.
These measures are invariant for the $\xbi$ process, for all $\phi\in [0,1]$. 

%%%%%%%%%%%%%%%%%%%%%%%%%%%%%%%%%%%%
\subsection{Variational formulation of  \texorpdfstring{$D_{\rm s}$}{Ds}}
%%%%%%%%%%%%%%%%%%%%%%%%%%%%%%%%%%%%

The self-diffusion coefficient has a variational characerisation in terms of the $\xbi$ process \cite{Sophn}. In the current notation, it is expressed as
\begin{multline}
	 \hat{D}^{kk}_s(\phi)  = \frac{1}{2} \inf_f  \EE_{\mu_\phi} \bigg[ \sum_{e \in B_0} [1- \xbi(e)] [ e \cdot e_k  +f (\xbi) - f( \tau_{e} \xbi) ]^2 
	 \\ + \frac{1}{2} \sum_{\substack{x,y\in \ZZ^d_*\\ \vert x-y\vert =1}}  [ f( \xbi^{x y} ) -f (\xbi) ]^2 \bigg],
\end{multline}
where the infimum is over cylinder functions $f$.
For any two such cylinder functions $f_1,f_2$, define the inner product
\beq
\langle f_1,f_2 \rangle = \sum_{\xbi \in \{0,1\}^\zdstar} f_1(\xbi) f_2(\xbi).
\eeq
Then a computation shows that\footnote{%
Equation \eqref{equ:var-int} corresponds to equation (5.1) of \cite{lamdim}.
We note two key differences. The first one is to do with the convention used for the diffusion coefficient: in \cite{lamdim} they define $D_s$ such that the limiting Brownian motion is $\sqrt{D_s} {\mathrm d} W$ and  the jump rates $\lambda_\s$ via a generic symmetric law $p(v)$ with finite range. Instead we have $\sqrt{2D_s}{\mathrm d}W$ and rates $p(v) = 1_{\vert v \vert=1}$. 
Second, we note there is a missing factor of two in front of the $\alpha(1-\alpha)$ term in equation (5.1) of \cite{lamdim} (where $\alpha$ corresponds to our $\phi$), cf. equation (2.7) of \cite{finitelatticediff} for the corrected version. 
}
\beq
 \hat{D}^{kk}_s(\phi) =  (1-\phi) - \phi(1-\phi) \sup_f \left[ 2\langle f,g_k\rangle + \langle {\cal L}^{\rm t} f, f \rangle ],
 \right]
 \label{equ:var-int}
\eeq
where $g_k$ is a cylinder function given by
\begin{align}\label{equ:def-gk}
\begin{aligned}
g_k(\xbi) & = \frac{1}{\sqrt{\phi(1-\phi)}} \sum_{e\in B_0} (e\cdot e_k) [ 1 - \xbi(e) ] \\
&= \frac{1 }{\sqrt{\phi(1-\phi)}} \left[ \xbi(-e_k) - \xbi(e_k) \right]   .
\end{aligned}
\end{align}

%%%%%%%%%%%%%%%%%%%%%%%%%%%%%%%%%%%%
\subsection{A basis for the space of cylinder functions, and the corresponding generator}
%%%%%%%%%%%%%%%%%%%%%%%%%%%%%%%%%%%%

\newcommand{\siteset}{{\cal S}}

We consider a basis for these functions to perform the supremum over cylinder functions $f$. A similar approach was taken~\cite{lamdim}, but this differs in the details.  
 Noting that, by definition, $f$ depends on $\xbi$ through a finite number of sites,  it is possible to represent a generic cylinder function as
\beq
f(\xbi) = a_0 + \sum_{x\in S_1} a_1(x) \xi(x) + \sum_{(x,y)\in S_2} a_2(x,y) \xi(x)\xi(y) + \cdots,
\label{equ:f-basis-bad}
\eeq
where each $S_n\subset (\zdstar)^n$ is finite, $a_n:S_n\to\mathbb{R}$, and every $f$ has a representation with a finite number $k\ge 0$ of sums on the right-hand side. The function $f$ is said to have degree $k$. [For example, the one-particle density $P_\s(x)$ is the expectation of a degree one function, while the two-particle density $P_{\s,\s'}(x,y)$ is the expectation of a degree two function, see \eqref{equ:pij}.]

For each integer $n \geq 0$, let $\siteset_n$ denote all the subsets of $\ZZ^d_*$ with $n$ points. Define $\siteset = \bigcup_{n\geq 0} \siteset_n$. Then for each $A \in \siteset$ define 
\begin{equation}
	\Psi_A (\phi, \xbi)  = \prod _{x \in A} \frac{\xbi(x)-\phi}{\sqrt{\phi(1-\phi)}},
\end{equation}
and $\Psi_\emptyset =1$.  Note that $\Psi_A(\phi,\cdot)$ is a cylinder function with the orthonormality relation 
\beq
\mathbb{E}_{\mu_\phi} \big[ \big\langle \Psi_A(\phi,\cdot) \Psi_B(\phi,\cdot) \big\rangle \big] = \delta_{A,B},
\label{equ:ortho}
\eeq
where $\delta_{A,B}=1$ if the sets $A$ and $B$ are equal, and zero otherwise.
Moreover, an arbitrary cylinder function can be expressed as
\beq
	f(\xbi) = \sum_{A \in \siteset} \Psi_A(\phi,\xbi)  \mathpzc{f}(A),
\label{equ:f-rep}
\eeq
with $\mathpzc{f}:\siteset \rightarrow \RR$.  This representation is more abstract than \eqref{equ:f-basis-bad} because $\mathpzc{f}$ is a function whose domain is a set of sets, but this abstraction makes it more convenient for later analysis.  
If $f$ is of degree $\leq k$, then $\mathpzc{f}$ is supported in $\bigcup_{n=0}^k \siteset_n$.
For two functions $f_1,f_2$ represented as (\ref{equ:f-rep}), define
\beq
\langle \mathpzc{f}_1, \mathpzc{f}_2 \rangle = \sum_{A \in \siteset} \mathpzc{f}_1(A) \mathpzc{f}_2(A) ,
\eeq
so by (\ref{equ:ortho})
\beq
\mathbb{E}_{\mu_\phi} \big[ \big\langle f_1,f_2 \big\rangle \big] = \langle \mathpzc{f}_1, \mathpzc{f}_2 \rangle.
\eeq

For any given $\phi$, the functions $f$ and $\mathpzc{f}$ are in one-to-one correspondence.
It follows that when the generator ${\cal L}^{\rm t}$ acts on $f$ to produce a new cylinder function ${\cal L}^{\rm t}f$, there is a corresponding operator $\mathpzc{L}_\phi$ such that
\beq
	{\cal L}^{\rm t}f(\xbi) = \sum_{A \in \siteset} \Psi_A(\phi,\xbi) \mathpzc{L}_\phi\mathpzc{f}(A).
\eeq
The operator $\mathpzc{L}_\phi$ can be constructed as follows. First for a  finite subset $A \subset \ZZ^d_*$ and $x,y \in \ZZ^d_*$ define the set operations
\begin{align*}
		A^{x,y} &= \begin{cases}
					(A \setminus \{x\}) \cup \{y\} & \text{if } x \in A, y \not\in A, \\
					(A \setminus \{y\}) \cup \{x\} & \text{if } y \in A, x \not\in A, \\
					A & \text{otherwise},
				  \end{cases} 
\end{align*}
and
\begin{align*}
		S_x A &= \begin{cases}
					A-x & \text{if } x \not\in A, \\
					(A \setminus \{x\}) \cup \{-x\} & \text{if } x \in A ,
				\end{cases}  
\end{align*}
where $A-x$ denotes the set $\{ y \in \zdstar : y+x\in A\}$.  
Then $\mathpzc{L}_\phi $ can be written as, 
\begin{subequations} \label{Lphi}
	\begin{equation} \label{Lphi_def}
		\mathpzc{L}_\phi = \mathpzc{L}_0+\phi \mathpzc{L}^1_\tau+(1-\phi)\mathpzc{L}^2_\tau+\sqrt{\phi(1-\phi)} (\mathpzc{L}^+_\tau+\mathpzc{L}^-_\tau)   ,
\end{equation}
where 
\begin{align}
\begin{aligned}
\label{equ:Loperator}
		     (\mathpzc{L}_0 \mathpzc{f}) (A) &= \frac{1}{2} 
		     \sum_{x,y\in\zdstar} \sum_{x-y\in B_0}
		     [ \mathpzc{f}( A^{x,y} ) -\mathpzc{f} (A) ],\\
		(\mathpzc{L}^1_\tau \mathpzc{f}) (A) &= \sum_{x \in A \cap B_0}  [\mathpzc{f}( S_x A ) -\mathpzc{f} (A) ],\\
		(\mathpzc{L}^2_\tau \mathpzc{f}) (A) &= \sum_{x \in \bar{A} \cap B_0} [\mathpzc{f}( S_x A ) -\mathpzc{f} (A)],	\\
		(\mathpzc{L}^+_\tau \mathpzc{f}) (A) &= \sum_{x \in A \cap B_0} [\mathpzc{f}( A \setminus \{x\} ) -\mathpzc{f} (S_x A \setminus \{-x \} ) ],	\\
		(\mathpzc{L}^-_\tau \mathpzc{f}) (A) &= \sum_{x \in \bar{A} \cap B_0} [ \mathpzc{f}( A \cup \{x\} ) -\mathpzc{f} (S_x A \cup \{-x \} ) ]	 ,
\end{aligned}
\end{align}
\end{subequations}
in which $\bar{A}=\zdstar \setminus A$.

%%%%%%%%%%%%%%%%%%%%%%%%%%%%%%%%%%%%
\subsection{Dependence of self-diffusion constant on  \texorpdfstring{$\phi$}{psi} }
\label{appendix: dependence}
%%%%%%%%%%%%%%%%%%%%%%%%%%%%%%%%%%%%

So far, our discussion follows~\cite{lamdim}.  In this subsection we follow the suggestion of Remark 5.3 of that work, to compute the dependence of the self-diffusion constant on $\phi$. Define a variable $\theta \in [0,\pi/2]$ such that 
\beq
\phi = \sin^2 \theta, \qquad \sqrt{4\phi(1-\phi)} = {\sin 2\theta},
\eeq
and $\mathpzc{L}(\theta) = \mathpzc{L}_\phi$. 
Note that, since $g_k$ in \eqref{equ:def-gk} is a cylinder function, it may be represented as 
\beq
g_k(\xbi) = \sum_{A \in \siteset} \Psi_A(\phi,\xbi)  \mathpzc{g}_k(A),
\eeq 
with
\begin{equation}
	\mathpzc{g}_k( A ) = 
	\begin{cases}
		- \v   \cdot e_k & \text{if}~ A=\{\v\} ~\text{with}~ \vert \v\vert  =1, \\
		0 & \text{otherwise}.
	\end{cases}
\end{equation}
Note that this $g_k$ is of degree one (because $\mathpzc{g}_k$ is supported on sets with only one element).

A crucial result of~\cite{lamdim} is (see their equation (5.3))
\beq
\sup_f \mathbb{E}\left[ 2\langle f,g_k\rangle + \langle {\cal L}^{\rm t} f, f \rangle  \right]
 = \lim_{\lambda\to0} \langle  \mathpzc{g}_k, \mathpzc{u}_{k} \rangle,
 \label{equ:james-9}
\eeq
where $\mathpzc{u}_{k} = \mathpzc{u}_{k}(\lambda,\theta; A)$ is defined to be the solution to
\beq \label{uk_eq}
		\lambda \mathpzc{u}_k(\lambda,\theta; A) - \mathpzc{L}(\theta)\mathpzc{u}_k(\lambda,\theta; A)= \mathpzc{g}_k(A).
\eeq

Recalling (\ref{equ:var-int}), the behavior of $D_s$ close to $\phi=0$ and $\phi=1$ is available by computing the right-hand side of (\ref{equ:james-9}) at $\theta=0$ and $\theta=\pi/2$, respectively. In \cite{lamdim} it is shown that the limit $\lambda\to0$ in \eqref{uk_eq} is well-behaved for all $\theta$, including the end points $\theta =0,\pi/2$. Therefore, we are left to compute the solution of
\beq
		- \mathpzc{L}(\theta)\mathpzc{u}_k(0,\theta; A)= \mathpzc{g}_k (A).
		\label{equ:L-u0-g}
\eeq
The right-hand side has degree one. From \eqref{equ:Loperator} one sees that for $\theta=0,\pi/2$ then $\mathpzc{L}(\theta)\mathpzc{f}$ has the same degree as $ \mathpzc{f}$.  Hence, in these cases, $\mathpzc{u}_k(0,\theta;A)$ is of degree one, so there is a function $\chi_{k,0}:\zdstar\to\mathbb{R}$ such that
\begin{equation*}
	\mathpzc{u}_k(0,0; A ) = 
	\begin{cases}
		\chi_{k,0}(\v) & \text{if}~ A=\{\v\} ~\text{with}~ \v\in\zdstar,  \\
		0 & \text{otherwise},
	\end{cases}
\end{equation*}
and similarly for $\mathpzc{u}_k(0,\pi/2; A )$ with corresponding function $\chi_{k,\pi/2}$.
To solve (\ref{equ:L-u0-g}), we consider the action of $\mathpzc{L}(t)$ on these $\mathpzc{u}_k$ functions. Combining \eqref{Lphi} and \eqref{equ:L-u0-g} for $\theta=0,\pi/2$ yields 
\begin{align}
\begin{aligned}
\label{equ:chi-eq}
	2\lap \chi_{k,0  }(\v)
	&=  [2\chi_{k,\pi/2}(0) - 2\chi_{k,\pi/2}(\v) + \v\cdot e_k] \id_{\{ \vert \v\vert =1  \} } ,
	\\
	 \lap \chi_{k,\pi/2}(\v)  
	&=  [\chi_{k,\pi/2}(0) - \chi_{k,\pi/2}(-\v) + \v\cdot e_k] \id_{\{ \vert \v\vert =1  \} },
\end{aligned}
\end{align}
where we recall that $\lap$ is the standard Laplacian with unit spacing. Comparing with the inner problem \eqref{inner1}, the factor of two corresponds to $D_\s+D_{\s'}$ which in this section simplifies to two. As in Subsection \ref{sec:inout}, these problems are solved by a multiple of the auxiliary function $\psi$ solving \eqref{aux_problem} and satisfying \eqref{prop_green}.  Recalling from Appendix~\ref{app:psi} that $\psi_j (0) = 0,  \psi_j(\pm e_k) = \pm \beta \delta_{jk}$ and using \eqref{alpha_def}, it is easy to verify 
\begin{align}
\begin{aligned}
\label{equ:chi-sol}
	\chi_{k,0}(\v)  &=\frac{1}{2(1+\beta)}\psi_k(\v)= \frac{1+\alpha}{2}\psi_k(\v)
	\\
	\chi_{k,\pi/2}(\v) &= \frac{1}{(1-\beta)}\psi_k(\v) = \frac{1+\alpha}{1+2\alpha}\psi_k(\v)   .
\end{aligned}
\end{align}

To compute the first order correction to $D_s$, we evaluate \eqref{equ:var-int}. In particular \eqref{equ:james-9} corresponds to the contribution missed by the mean-field approximation. Taylor expanding \eqref{equ:james-9} about $\theta = 0$ and combining with \eqref{equ:var-int} yields
\beq
 D_s(\phi) =  (1-\phi) - \phi(1-\phi) \left( \langle  \mathpzc{g}_k, \mathpzc{u}_{k}(0, 0 ) \rangle + \theta \partial_ \theta  \langle  \mathpzc{g}_k, \mathpzc{u}_{k}(0,0) \rangle + O(\theta^2) \right),
 \label{equ:Dstaylor}
\eeq
and similarly for $\theta = \pi /2$. Combining with \eqref{equ:chi-sol}, the expansions \eqref{equ:Dexpan linear} follows. The $-\phi \langle  \mathpzc{g}_k, \mathpzc{u}_{k}(0, 0 ) \rangle = -\alpha \phi$ term, corresponds to the part of the interaction term, $- \alpha p_{\s'}\nabla p_{\s} $, resulting from the  $\psi$ term in the inner solution $P_\text{in}$.

%%%%%%%%%%%%%%%%%%%%%%%%%%
\subsection{Second-order expansion of \texorpdfstring{$D_s$}{Ds} } \label{sec:second_order_recursive}
%%%%%%%%%%%%%%%%%%%%%%%%%%

Recalling (\ref{equ:Dstaylor}), higher order terms in the expansion of $D_s$ are also available by computing Taylor expansions of (\ref{equ:james-9}) about $\theta =0$ or $\theta =\pi/2$.
In \cite{lamdim} they prove uniform convergence of $\mathpzc{u}_{k}(\lambda,t)$ and its derivatives, and therefore
\beq
\frac{d^n}{d \theta ^n} \lim_{\lambda\to0} \langle  \mathpzc{g}_k, \mathpzc{u}_{k}(\lambda,\theta ) \rangle = 
\lim_{\lambda\to0} \langle  \mathpzc{g}_k, \partial^n_\theta
\mathpzc{u}_{k}(\lambda,\theta ) \rangle   .
\eeq
By differentiating \eqref{equ:L-u0-g}, we can recursively solve for $ \partial^n_\theta \mathpzc{u}$, and therefore compute $D_s$ to an arbitrary order. 

In order to compare this method with matched asymptotics at $O(\phi^2)$, we compute the next coefficient about $\phi = 0$. To do this we require the order $\theta^2 \sim \phi$ terms in the Taylor expansion of $\mathpzc{u}$ about zero. As $D_s$ is differentiable in $\phi$, we do not expect an $O(\theta)$ contribution from $\p_\theta\mathpzc{u}$, however its computation is still required to solve for $\p_\theta^2 \mathpzc{u}$. 
Differentiating \eqref{equ:L-u0-g} yields
\begin{align*}
 		- \mathpzc{L}(0) \partial_\theta\mathpzc{u}(0,0;A) &= \mathpzc{L}'(0)\mathpzc{u}(0,0;A),
		\\
		- \mathpzc{L}(0) \partial_\theta^2 \mathpzc{u}(0,0;A) &=  
		2\mathpzc{L}'(0)\partial_\theta \mathpzc{u} 
		+ \mathpzc{L}''(0)\mathpzc{u}(0,0;A)   .
\end{align*} 
Evaluating $\mathpzc{L}(\theta)$ and its first and second derivatives $\mathpzc{L}'(\theta)$ and $\mathpzc{L}''(\theta)$  at $\theta = 0$ using \eqref{Lphi_def} yields:
\begin{align}
\label{equ:u2}
 		\p_\theta \mathpzc{u} &= - (\mathpzc{L}_0+\mathpzc{L}^2_\tau)^{-1} 
 		(
 		\mathpzc{L}^-_\tau +\mathpzc{L}^+_\tau 
 		)
		\mathpzc{u}, 		\\
		\p_\theta^2 \mathpzc{u} &= - 2(\mathpzc{L}_0+\mathpzc{L}^2_\tau)^{-1}		(
		\mathpzc{L}^-_\tau +\mathpzc{L}^+_\tau 
		)
		\mathpzc{u}^{(1)}
		+
		(
		1-2 \alpha
 		) 2\mathpzc{u}  . \nonumber
\end{align}
These equations fully determine $\p_\theta \mathpzc{u}$ and $\p_\theta^2 \mathpzc{u}$ but we do not solve for these explicitly. It is notable that $\mathpzc{L}^+_\tau$ increases the degree of a function by one, $\mathpzc{L}^-_\tau$ decreases the degree of a function by one and $(\mathpzc{L}_0+\mathpzc{L}^2_\tau)$ does not change the degree of a function. Hence $\p_\theta \mathpzc{u}$ will be a degree two function with no degree one component and $\p_\theta^2 \mathpzc{u}$ will be degree three with a degree one component. As $\mathpzc{g}_k$ is a degree one function $\langle \mathpzc{g}_k, \p_\theta \mathpzc{u}(0,0) \rangle =0 $, as expected. The next non-zero term in \eqref{equ:Dstaylor} will be $\frac{1}{2}\phi^2 \langle \mathpzc{g}_k, \p^2_\theta \mathpzc{u}(0,0) \rangle = \p_\theta^2\mathpzc{u}_k( \{e_k \})$. Collecting $O(\phi^2)$ terms in \eqref{equ:Dstaylor}, the next term in the expansion of $D_s(\phi)$ is $[\p_\theta^2\mathpzc{u}_k( \{e_k \})-2\mathpzc{u}_k( \{e_k \})]\phi^2$. In the following section we show that matched asymptotics gives the same result.

%%%%%%%%%%%%%%%%%%%%%%%%%%%%%%%%%%%%
\section{Second-order matched asymptotics}
\label{appendix:matched second}
%%%%%%%%%%%%%%%%%%%%%%%%%%%%%%%%%%%%

In this appendix, we show the calculation via matched asymptotics of the order $\phi^2$ contribution to equation \eqref{equ:asymp soln} for $\phi\ll 1$, and how it is consistent with the second-order expansion of $D_s$ in Subsection~\ref{sec:second_order_recursive} using the rigorous recursive method. This requires evaluating the $O(\phi)$ terms in equation \eqref{equ:two particle}, which as discussed correspond to three-particle interactions.  
We show the derivation for the simplest case of equal diffusivities and no drifts, $D_r = D_b = 1, V_r = V_b \equiv 0$, since this is also the case where the recursive method of Appendix \ref{appendix:Ds} applies. 

We recall the starting point: the one-particle density $P_\s$ satisfies \eqref{integrated_eq} exactly, with interaction terms $\mathcal{E}_{\s,\s}$ and $\mathcal{E}_{\s,\os}$ in \eqref{equ:inter-explicit} depending on the two-particle densities $P_{\s,\s}$ and $P_{\s,\os}$, respectively. In Subsection \ref{sec:Asymp} we have computed the leading-order contributions in $\mathcal{E}_{\s,\s}$ and $\mathcal{E}_{\s,\os}$, which were obtained by considering the equation satisfied by the two-particle density $P_{\s,\s'}$ \eqref{equ:two particle} to leading order and approximating its solution via inner and outer asymptotic expansions. In order to obtain the next asymptotic term in the equation for $P_\s$, we need to consider \eqref{equ:two particle} to $O(\phi)$:
\begin{subequations} \label{two_part_3}
\begin{align} \label{equ:two particle plus}
\begin{aligned}
	\dot P_{\s,\s'}(x,y,t) & =
	\sum_{ \substack{z \in \HH \\ z \neq y} }
	 \left[
	\lambda_{\s}  (z,x) P_{\s,\s'}(z,y,t) -\lambda_{\s}  (x,z) P_{\s,\s'}(x,y,t)
	\right]
\\
	& \phantom{=} + \sum_{ \substack{z \in \HH \\ z \neq x} }
	 \left[
	\lambda_{\s'}  (z,y) P_{\s,\s'}(x,z,t) -\lambda_{\s'}  (y,z) P_{\s,\s'}(x,y,t)
	\right] \\
& \phantom{=} + \mathcal{E}^{\rm int}_{\s,\s'} (x,y,t)  ,
\end{aligned}
\end{align}
where the three-particle interaction term $\mathcal{E}^{\rm int}_{\s,\s'}$ is defined analogously to the two-particle interaction \eqref{equ:inter} as 
\begin{align}
\begin{aligned}
	\label{3-part-interaction}
	\mathcal{E}^{\rm int}_{\s,\s'} (x,y,t) &= h^d (N_\s - 1 - \id_{\s'= \s}) \mathcal{E}_{\s,\s',\s}(x,y,t) \\
	& \phantom{=} + h^d (N_\os - \id_{\s'= \os}) \mathcal{E}_{\s,\s',\os}(x,y,t),
\end{aligned}
\end{align}
and
 \begin{align}
 \begin{aligned}
	\mathcal{E}_{\s,\s',\s''}(x,y,t) &= 
	\frac{1}{h^2} 
	\sum_{ \substack{
	 z \in \Omega \\ 
	 z \neq \xx \\
	\vert z-x\vert  = h
	}}
	\Big[ 
	 P_{\s,\s',\s''} (x, \xx , z, t)
	-P_{\s,\s',\s''} (z, \xx , x, t)
	\Big] 
	\\
	& \quad +
	\frac{1}{h^2}
	\sum_{ \substack{
	 z \in \Omega \\ 
	 z \neq x \\
	\vert z-\xx\vert  = h
	}}
	\Big[ 
	 P_{\s,\s',\s''} (x ,\xx,z ,t)
	-P_{\s,\s',\s''} (x ,z ,\xx,t)
	\Big] .
\label{equ:3int} 
\end{aligned}
\end{align}
\end{subequations}
Thus we see that the three-particle interactions in a two-species mixture involve the three-particle densities $P_{\s,\s,\s}(\x,t), P_{\s,\s,\os}(\x,t)$ and $P_{\s,\os,\os}(\x,t)$ where $\x = (x,\xx,\xxx) \in \Omega^3_h$ and $\Omega_h^3 = \Omega^3 \setminus \{x = y, x=z, y=z\}$ 
\beq
P_{\s,\s',\s''}(\x, t) = \frac{\EE [ \eta^\s_t(x)\eta^{\s'}_t(\xx) \eta^{\s''}_t(\xxx)]}{N_\s \tw{N}_{\s'} \tw{N}_{\s''} h^{3d} }  , \text{ for } \x \in \Omega_h^3, %x \neq y, ~y\neq z,~ z \neq x,
\eeq
where $\tw{N}_{\s'} = N_{\s'} -\id_{\s'= \s}$ and $\tw{N}_{\s''} = N_{\s''} -1 -\id_{\s''= \s}$.

Our goal in this Appendix is to obtain an asymptotic approximation for small $\phi$ of \eqref{equ:3int} using matched asymptotic expansions. We proceed with the same method as in Subsection \ref{sec:Asymp}, namely, to consider the `asymptotically closed' (for $\phi\ll 1$) equation for the three-particle density $P_{\s,\s',\s''}(\x, t)$ and divide its domain of definition into regions depending on whether three, two or no particles are close to one another. 

We make two simplifications specific to the case $V_r = V_b =0$, but emphasise that the method easily extends to the general case. As $V_r = V_b =0$, it follows by the symmetry of $P_{\s, \s}$ that ${\cal E}_{\s, \s} = 0$ and therefore we only require $P_{\s, \os}$ to evaluate ${\cal E}^{\rm int}$. As a third particle must be either $\s$ or $\os$ without loss of generality we solve for $P_{\s, \os, \os}$, and by permuting species can evaluate ${\cal E}^{\rm int}_{\s, \os}$.

%%%%%%%%%%%%%%%%%%%%%%%
\subsection{Matched asymptotics expansion for  \texorpdfstring{$P_{\s,\os,\os}$}{Pss's''} }
%%%%%%%%%%%%%%%%%%%%%%%

Setting $f(\eta) = \eta^\s_t(x)\eta^{\s'}_t(\xx) \eta^{\s''}_t(\xxx)$ in \eqref{Generator}, it follows by similar calculation to Section \ref{sec:Prelim} that
\begin{align}
\begin{aligned}
	\label{3lMargODE}
	h^2\dot P_{\s,\s',\s''} (\x , t)
	&= \sum_{ \substack{ w \in \Omega \\ w \neq \xx, \xxx \\
	\vert w-x\vert  = h
	}} \left[ P_{\s,\s',\s''}(w,\xx,\xxx,t)- P_{\s,\s',\s''}(\x,t) \right] \\
	& \phantom{=} \sum_{ \substack{ w \in \Omega \\ w \neq x, \xxx \\
	\vert w-\xx\vert  = h
	}} \left[ P_{\s,\s',\s''}(x,w,\xxx,t)- P_{\s,\s',\s''}(\x,t) \right] \\
	& \phantom{=}  \sum_{ \substack{ w \in \Omega \\ w \neq x, \xx \\
	\vert w-\xxx\vert  = h}} \left[ P_{\s,\s',\s''}(x,\xx,w,t)- P_{\s,\s',\s''}(\x,t) \right] + O(\phi) ,
\end{aligned}
\end{align}
for $\x\in \Omega_h^3$. We proceed by introducing three regions of $\Omega_h^3$: an inner region where the three particles are within $O(h)$ of each other; an intermediate region where two particles are close and the third one is far; and an other region where the three particles are far from each other. Consequently, we define 
\beq
P_{\s,\s',\s''}(\x,t) = \begin{cases} 
\tw{P}_{\rm out}(x,\xx,\xxx, t),
& \vert x-\xx\vert, \vert x-\xxx\vert, \vert \xx-\xxx\vert  \gg h, 
\\
\tw{P}_{\rm mid}\big(x,\frac{\xx-x}{h},\xxx,t\big), &  \vert x-\xx\vert  \sim h, \vert x-\xxx\vert \gg h, \\ 
\tw{P}_{\rm in}\big(x,\frac{\xx-x}{h},\frac{\xxx-x}{h},t\big), & \vert x-\xxx\vert  \sim h, \vert x-\xx\vert \sim h.
\end{cases}
\label{equ:out-in 2nd}
\eeq
As before, we omit the dependence on $\s,\s',\s''$ of $\tw{P}_{\rm out}$, $\tw{P}_{\rm mid}$ and $\tw{P}_{\rm in}$  for ease of notation.
Following the procedure in Subsection~\ref{sec:Asymp}, we keep the boundary layer variables in $\zdstar$ (discrete) while taking the rest to be continuous.

\paragraph*{Outer region}

In the outer region, analogously to \eqref{equ:two outer}, we have
\begin{align}
\label{equ:two outer 3}
\begin{aligned}
	{\dot {\tilde P}}_{\text{out}}(\x, t) 
& =  \mathcal{L}^*_\s P_{\text{out}}(\cdot,y,z,t) + \mathcal{L}^*_{\s'} P_{\text{out}}(x,\cdot,z,t) \\
& \phantom{= } + \mathcal{L}^*_{\s''} P_{\text{out}}(x,y,\cdot,t) +O(\phi)  ,
\end{aligned}
\end{align}
with solution
\begin{equation}
\label{equ:three outer sol}
\tw{P}_{\text{out}} (\x,t) = p_{\s} (x,t)p_{\s'} (\xx,t)p_{\s''}(\xxx,t) + h \tw{P}_{\text{out}}^{(1)} (\x,t) + \cdots    .
\end{equation}

\paragraph*{Intermediate region}

In the intermediate region in which $\vert x -\xx \vert \sim h$ and $\vert x -\xxx \vert \gg h$, we introduce the intermediate coordinates
\begin{align*}
	x = \xmid, \qquad \xx = \xmid + h\ymid, \qquad \xxx = \zmid,
\end{align*}
and write $P(\x,t) = \tilde P_\text{mid}(\xmid,\ymid,\zmid,t)$.
By identical calculation to the first-order inner problem (Subsec. \ref{sec:Asymp}) and imposing the matching condition to the outer region,
$$
\tilde P_\text{mid}(\xmid,\ymid,\zmid,t) \sim \tw{P}_{\text{out}}(\xmid, \xmid+h\ymid,\zmid,t), \qquad \text{as } \vert \ymid \vert \to \infty.
$$
it follows that for $\zmid \in \Omega$
\begin{equation}
    \tw{P}_{\text{mid}}(\xmid,\ymid,\zmid,t) = P_\text{in}(\xmid, \ymid, t) P_{\s''} (\zmid,t)   , \label{equ:meanfieldish}
\end{equation}
where $P_\text{in}(\hat x, \hat y, t)$ satisfies the two-particle inner problem \eqref{inner_problem} and is therefore given by \eqref{equ:inner soln}. For symmetric and equal rates,  $\tw{P}_{\text{mid}}$ reduces to
\begin{subequations} \label{inner_simplified}
	\begin{align} \label{pin simplified}
P_\text{in}(\xmid, \ymid, t) = \ & p_s(\xmid,t)p_{s'}(\xmid,t) p_{\s''}(\xmid,t) \\
&+ h \left[ p_{\s''} \bfA_{\s,\s'} \cdot \mathpzc{u}( \{ \ymid  \} ) + p_{\s''} \bfB_{\s,\s'} \cdot \ymid + \bfC_{\s,\s'} \right] + O(h^2,\phi), \nonumber
\end{align}
where
\begin{align} \label{coeffs_inner_general}
\begin{aligned}
\bfA_{\s,\s'}(\xmid,t) &=  (p_{\s'} \nabla p_\s - p_\s \nabla p_{\s'}),\qquad
\bfB_{\s,\s'}(\xmid,t) =   p_\s \nabla p_{\s'}, \\
\bfC_{\s,\s'}(\xmid,t) &= \tw{P}_{\text{out}}^{(1)} (\xmid,\xmid,\zmid,t).
\end{aligned}
\end{align}
\end{subequations}
and as introduced in Sec. \ref{appendix:Ds}, $\mathpzc{u}( \{ \ymid  \} )= \mathpzc{u}(0,0; \{ \ymid  \} ) = \frac{{\bgfpsi}(\ymid)}{ 2 (1 + \beta)}$

\paragraph*{Inner region}

As discussed, we now consider the case $\s',\s'' = \os$, which is sufficient to deduce ${\cal E}^{\rm int}$ in the case of no potential. This will allow us to solve the inner problem in terms of the function $\mathpzc{u}$, from Sec. \ref{appendix: dependence}, therefore making comparison simpler.  
Now consider the inner region where we use the coordinates: 
\begin{align*}
	x  = \xin, && \xx = \xin + h\yin ,&& \xxx = \xin +h \zin  .
\end{align*}
As the second two species are of the same species, the $\yin, \zin$ coordinates are interchangeable in $\tw{P}_{\text{in}}$. 
We abuse notation and allow $\tw{P}_{\text{in}}(\xin, \yin,\zin , t)$ to also be denoted $\tw{P}_{\text{in}}(\xin, \{\yin,\zin \}, t)$, to allow the operators \eqref{equ:Loperator} to act on  $\tw{P}_{\text{in}}$ in the $\{\yin,\zin\}$ argument. 
In the inner coordinates, \eqref{3lMargODE} becomes:
\begin{subequations}
\label{inner3}
\begin{align} 
	\dot{\tw{P}}_{\text{in}}
	= h^{-2}( \mathpzc{L}_0+\mathpzc{L}^2_\tau)  \tw{P}_{\text{in}} 
	-h^{-1} (
 		\yin \id_{ \{ \vert \yin \vert = 1 \} } +\zin \id_{ \{ \vert \zin \vert = 1 \}} 
 		) \cdot \nabla_\xin \tw{P}_{\text{in}}
 +O(h^{-1}\phi,1).
\end{align}
The inner problem is complemented by the matching conditions with the intermediate densities \eqref{equ:meanfieldish}
\begin{align}
	\tw{P}^{(1)}_{\text{in}} &\sim P_\text{in}(\xin, \zin, t) P_{\os} (\xin + h \yin,t), \qquad \text{as } \vert \yin \vert \to \infty,\\
	\tw{P}^{(1)}_{\text{in}} &\sim P_\text{in}(\xin, \yin, t) P_{\os} (\xin + h \zin,t), \qquad \text{as } \vert \zin \vert \to \infty.
\end{align}
\end{subequations}
We look for a solution to \eqref{inner3} of the form $
		\tw{P}_{\text{in}} (\xin, \yin, \zin, t) \sim	\tw{P}^{(0)}_{\text{in}} (\xin, \yin, \zin, t)+h\tw{P}^{(1)}_{\text{in}} (\xin, \yin, \zin, t) + \cdots$. 
At $O(h^{-2})$ we find
\begin{subequations} \label{inner3_0}
\begin{alignat}{3}
	0 &= ( \mathpzc{L}_0+\mathpzc{L}^2_\tau) \tw{P}^{(0)}_{\text{in}}, &\qquad &(\yin, \zin) \in \zdstart,\\
	\tw{P}^{(0)}_{\text{in}} &\sim P_\text{in}^{(0)}(\xin, \zin, t) p_{\os} (\xin,t), & & 
	\yin \to \infty,\\
	\tw{P}^{(0)}_{\text{in}} &\sim P_\text{in}^{(0)}(\xin, \yin, t) p_{\os} (\xin,t), & & 
	\zin \to \infty,
\end{alignat}	
\end{subequations}
where $\zdstart = \ZZ^{2d} \setminus \{\yin = 0, \zin = 0, \yin = \zin\}$ and $\tw{P}^{(0)}_{\text{in}}$ are given in \eqref{inner_simplified}. By inspection we have that the solution to \eqref{inner3_0} is 
\begin{equation}
	\label{inner3_sol0}
	\tw{P}^{(0)}_{\text{in}} (\xin, \yin, \zin, t) = p_{\s} (\xin,t)p_{\os} (\xin,t)p_{\os}(\xin,t).
\end{equation}
The $O(h^{-1})$ of \eqref{inner3} is, using \eqref{inner3_sol0}, 
\begin{subequations} \label{inner3_1}
\begin{alignat}{3}
	0 &=  ( \mathpzc{L}_0+\mathpzc{L}^2_\tau) \tw{P}^{(1)}_\text{in} - (
 		\yin \id_{ \{ \vert \yin \vert = 1 \} } +\zin \id_{ \{ \vert \zin \vert = 1 \}}
 		) \cdot \nabla_\xin \tw{P}^{(0)}_{\text{in}} , &\qquad &(\yin, \zin) \in \zdstart,\\
	\tw{P}^{(1)}_{\text{in}} &\sim \left[\bfA_{\s, \os} \cdot {\bgfpsi}(\zin) + \bfB_{\s, \os} \cdot \zin\right] p_{\os}(\xin)   \nonumber & & \\
	 & \quad + \bfB_{\s,\os} \cdot \yin   p_{\os}(\xin) + \tilde P_\text{out}^{(1)}(\xin,\xin,\xin), & & 
	\yin \to \infty,\\
\tw{P}^{(1)}_{\text{in}} &\sim \left[\bfA_{\s,\os} \cdot {\bgfpsi}(\yin) + \bfB_{\s,\os} \cdot \yin\right] p_{\os}(\xin)   \nonumber & & \\
	 & \quad + \bfB_{\s, \os} \cdot \zin   p_{\os}(\xin) + \tilde P_\text{out}^{(1)}(\xin,\xin,\xin), & & 
	\zin \to \infty.
\end{alignat}	
\end{subequations}
We seek a solution to \eqref{inner3_1} of the form: 
\begin{align} \label{equ:2ndinnerexpansion}
\begin{aligned}
\tw{P}^{(1)}_{\text{in}}(\xin,\yin,\zin,t) & = \Psi(\xin,\yin,\zin,t ) + \left[\bfA_{\s, \os} \cdot \mathpzc{u}( \{ \zin \} ) + \bfB_{\s, \os} \cdot \zin\right] p_{\os}(\xin) \\
& \phantom{=} +\left[\bfA_{\s,\os} \cdot \mathpzc{u}( \{ \yin \} ) + \bfB_{\s,\os} \cdot \yin\right] p_{\os}(\xin) + 
 \tw{P}^{(1)}_{\mathrm{out}}(\xin, \xin, \xin, t),
		\end{aligned}
\end{align}
A simple computation shows that
\begin{align}
\nonumber	\left( \mathpzc{L}_0 +\mathpzc{L}^2_\tau \right) \Big\{ &\left[\bfA_{\s, \os} \cdot \mathpzc{u}( \{ \zin \} ) + \bfB_{\s, \os} \cdot \zin\right] p_{\os} \\
& +\left[\bfA_{\s,\os} \cdot \mathpzc{u}( \{ \yin \} ) + \bfB_{\s,\os} \cdot \yin\right] p_{\os} + \tw{P}^{(1)}_{\mathrm{out}}(\xin, \xin, \xin, t) \Big\} 
		\\
		= -\big( 
 		\yin &\id_{ \{ \vert \yin  \vert = 1 \} } +\zin \id_{ \{ \vert \zin \vert = 1 \}} 
 		\big) \cdot \nabla_\xin \tw{P}^{(0)}_{\text{in}} 
		-  p_{\os} \bfA_{\s,\os}\cdot ( \mathpzc{L}^-_\tau +\mathpzc{L}^+_\tau )\mathpzc{u}( \{ \yin, \zin \} ). \label{equ:expression1}
\end{align}
It follows that
\begin{alignat*}{3}
	\left( \mathpzc{L}_0 +\mathpzc{L}^2_\tau \right) \Psi(\xin,\yin,\zin,t ) & = p_{\os}(\xin,t) \bfA_{\s,\os} (\xin,t)  \cdot ( \mathpzc{L}^-_\tau +\mathpzc{L}^+_\tau )\mathpzc{u}( \{ \yin, \zin \} ) , &~~~ &(\yin, \zin) \in \zdstart,\\
	\Psi(\xin,\yin,\zin,t ) & \to 0, & & \yin, \zin \to \infty.
\end{alignat*}
It follows from \eqref{equ:u2} that $\Psi(\xin,\yin,\zin,t ) = - 2\p_\theta \mathpzc{u} (\{\yin,\zin \})\cdot p_{\os}(\xin,t) \bfA_{\s,\os} (\xin,t) $  and therefore
\begin{align}
\label{equ:3partinnersol}
\begin{aligned}
	\tw{P}^{(1)}_{\text{in}}(\xin,\yin,\zin,t ) & =  \tw{P}^{(1)}_{\mathrm{out}}(\xin, \xin, \xin, t)
		+( \yin + \zin)  \cdot \bfB_{\s,\os}(\xin,t) p_{\os}(\xin,t) \\
		& \phantom{=} + \left[ \mathpzc{u}( \{ \yin  \} )  + \mathpzc{u}( \{ \zin  \} ) - 2\p_\theta \mathpzc{u} (\{\yin,\zin \})  \right] \cdot  \bfA_{\s,\os} (\xin,t)p_{\os}(\xin,t) . 
		\end{aligned}
\end{align}

\paragraph*{Three-particle interaction term}

Here we evaluate the three particle interaction term, $\mathcal{E}_{\s,\os,\os}(x,y,t)$, to leading order in $\phi$ and $h$.

Consider the case $\vert x -z \vert \gg h$. In \eqref{equ:3int}, $P_{\s, \os \os}$ is evaluated when $\x$ is in the middle region and therefore we use the middle solution. In middle coordinates \eqref{equ:3int} becomes
\begin{align}
	\mathcal{E}_{\s, \os, \os}(\xmid,\zmid,t) &= 
	\frac{1}{h^2} 
	\sum_{
	 \ymid \in B_0 
	}
	\Big[ 
	 \tw{P}_{\text{mid}}(\xmid,\ymid,\zmid,t)
	-
	\tw{P}_{\text{mid}}(\xmid+h\ymid,-\ymid,\zmid,t)
	\Big],
\end{align}
where the second term dropped because $P_{\s,\os,\os}(x,y,z) =P_{\s,\os,\os}(x,z,y)$. 
Combining with \eqref{equ:meanfieldish} and \eqref{equ:inter-explicit} yeilds
\begin{equation}
	 \begin{aligned}
	\mathcal{E}_{\s,\os,\os}(\xmid,\zmid,t) &= p_{\os}(\zmid,t)\mathcal{E}_{\s,\os}(\xmid,t) +O(h, \phi).
\label{equ:3intmiddle}
\end{aligned}
\end{equation}
Now consider $\vert x-y \vert \sim \vert x-z \vert \sim h$. 
In \eqref{equ:3int}, $P_{\s, \os \os}$ is evaluated when $\x$ is in the inner region and therefore we use the inner solution.
In inner coordinates \eqref{equ:3int} becomes
\begin{align}
	\mathcal{E}_{\s, \os, \os}(\xin,\xin+h\yin,t) &= 
	\frac{1}{h^2} 
	\sum_{
	 \zin \in B_\yin
	}
	\Big[ 
	 \tw{P}_{\text{in}}(\xin,\yin,\zin,t)
	-
	\tw{P}_{\text{in}}(\xmid+h\zin,\yin,-\zin,t)
	\Big],
\end{align}
where the second term dropped because $P_{\s,\os,\os}(x,y,z) =P_{\s,\os,\os}(x,z,y)$. 
Combining with \eqref{equ:3partinnersol}
, \eqref{equ:Loperator} and \eqref{equ:chi-sol} yeilds
\begin{equation} \label{equ:3intin}
	\mathcal{E}_{\s,\os,\os}(\xin,\xin+h\yin,t) = h^{-1} p_{\os} \bfA_{\s,\os}  \cdot
 	\Big( (2-2 \alpha )
 	\yin \id_{\{ \vert \yin \vert  =1 \}}
 	- 2\mathpzc{L}^-_\tau  \p_\theta \mathpzc{u} (\{\yin \}) \Big).
\end{equation}
Finally noting $\mathcal{E}_{\s,\s',\s''}(x,y,t) = \mathcal{E}_{\s',\s,\s''}(y,x,t)$ we can deduce $\mathcal{E}_{\s,\os,\s} $ from \eqref{equ:3intmiddle} and \eqref{equ:3intin} using $\mathcal{E}_{\s,\os,\s}(x,y,t) = \mathcal{E}_{\os,\s,\s}(y,x,t)$. Hence we now have explicit expressions in terms of $p_\s$ and $p_{\os}$ for the three particle interaction term at leading order. In the following section we use this close equation \eqref{equ:two particle plus} to first order in $\phi$ and solve for $P_{\s,\os}$.

%%%%%%%%%%%%%%%%%%%%%%%%%%%%%%%%%%%%
\subsection{Equation for the one-particle density}
%%%%%%%%%%%%%%%%%%%%%%%%%%%%%%%%%%%%

Recall, that to close \eqref{equ:gen int eq} we require $P_{\s,\os}$ to evaluate ${\cal E}^\text{int}$. 
In the calculation in the main text we computed the first non-zero contribution of $\mathcal{E}_{\s,\s'}(x,t)$, which appears at $O(1)$. Here we want to go one order higher. 
Usually, the terms of interest could be the terms of order $h\phi$ or $\phi^2$ in the equation for $P_\s$. 
But here, as we take $h\to 0$ while keeping $\phi$ fixed, the former vanishes, and we are after the $O(\phi^2)$ contribution, coming from the leading-order contribution of an inner region with three particles. 
So the structure of $\mathcal{E}_{\s,\s'}(x,t)$ is as follows
$$
\mathcal{E}_{\s,\s'}(x,t) = \mathcal{E}_{\s,\s'}^{(0)}(x,t) + \phi \mathcal{E}_{\s,\s'}^{(1)}(x,t) + O(\phi^2, h).
$$
In the main text, we already computed $\mathcal{E}_{\s,\s'}(x,t)$ to the leading order, which remains unchanged. We set 
\begin{equation}
	\mathcal{E}_{\s,\s'}^{(0)} = (1+ \alpha ) \nabla \cdot (
	p_\s \nabla p_{\s'} 
	- p_{\s'} \nabla p_\s ) .
\end{equation}
As in the main text, we proceed by introducing two regions: the inner region where two particles are within $O(h)$ and an outer region where the two particles are far apart. And as before, 
\beq
P_{\s,\os}(x,y,t) = \begin{cases} 
P_{\rm out}(x,\xx, t),
& \vert x-\xx\vert \gg h, 
\\
P_{\rm in}\big(x,\frac{\xx-x}{h},t\big), & \vert x-\xx\vert \sim h.
\end{cases}
\label{equ:out-in 3rd}
\eeq
where we omit the dependence on $\s,\s'$ of $P_{\rm out}$ and $P_{\rm in}$  for ease of notation.
Following the procedure in Subsection~\ref{sec:Asymp}, we keep the boundary layer variables in $\zdstar$ (discrete) while taking the rest to be continuous.
We propose asymptotic expansions, but now in both $h$ and $\phi$ 
\begin{align}\label{equ:double expansion}
	P_{\rm out} &= P^{(0,0)}_\text{out} + h P^{(1,0)}_\text{out} + \phi P^{(0,1)}_\text{out} + h \phi P^{(1,1)}_\text{out}+ O(h^2,\phi^2), \\
	P_{\rm in} &= P^{(0,0)}_\text{in} + h P^{(1,0)}_\text{in} + \phi P^{(0,1)}_\text{in} + h \phi P^{(1,1)}_\text{in}+ O(h^2,\phi^2),
\end{align}
where terms are allowed to depend on $\phi$ through $p_\s$. We have already solved the leading order contributions in $\phi$ and therefore we set $P^{(n,0)}_\text{out} = P^{(n)}_\text{out}$, and  $P^{(n,0)}_\text{in} = P^{(n)}_\text{in}$ from the main text. In the following, we solve for $P_{\rm in}$ to $O(h\phi)$, allowing us to evaluate $\mathcal{E}_{\s,\os}^{(1)}$.

\paragraph*{Outer region}

In the outer region \eqref{equ:two particle plus} becomes, 
\begin{align}
    \label{equ: outer eq first}
    	\partial_t P_{\text{out}} &=
	  \mathcal{L}^*_\s P_{\text{out}}(\cdot,y,t) + \mathcal{L}^*_{\os} P_{\text{out}}(x,\cdot,t)
	\\
	& \quad +\phi_\s p_\s \mathcal{E}_{\os,\s} + 
	\phi_{\os} p_{\os} \mathcal{E}_{\s,\os}
+O(\phi^2, h\phi, h^d), \nonumber
\end{align}
using the independent walk or single particle adjoint generator \eqref{Generator1} (we write $\mathcal{L}^*_\s P_{\text{out}}(\cdot,y,t)$ to denote the operator acting on $P_{\text{out}}(x,y,t)$ as if it was a function of $x$ only for $y$ and $t$ fixed). Therefore, to first order in $\phi$, the evolution of $P_{\text{out}}$ corresponds to two independent $\sigma$ and $\os$ particles. That is, $P_{\text{out}}(x,y,t) = Q_\sigma(x,t) Q_{\os}(y,t)+O(\phi^2, h^d, h\phi)$ for some functions $Q_\sigma$ satisfying $\dot Q_\s = \mathcal{L}^*_\s Q_\sigma+\phi_{\os} \mathcal{E}_{\s,\os}$.
Using the normalisation condition, $ \sum_{x,y} h^{2d} P_{\s,\s'} = 1$, \eqref{equ:gen int eq} and \eqref{rel_P_p},
\begin{equation}
\label{equ: outer sol first}
P_{\text{out}} (x,\xx,t) = p_{\s} (x,t) p_{\os} (\xx,t)+ O(\phi^2,h)   .
\end{equation}
Therefore the $\phi$-dependence of $P^{(0,0)}_{\rm out}(x, \xx,t) = p_{\s} (x,t) p_{\os} (\xx,t)$, contained in $p_\s$ and $p_\os$, is correct to $O(\phi)$. Hence no correction is needed at $O(\phi)$ and we set $P^{(0,1)}_{\rm out} = 0$. 

\paragraph*{Inner region}

In the inner region the interaction term does not contribute at $O(h^{-2})$ and therefore it follows by identical calculation to Section \ref{sec:Asymp} that at $O(\phi)$
\begin{equation}
    P_{\text{in}}^{(0,1)}(\u , \v ,t) = P_{\text{out}}^{(0,1)} (\u , \u , t) = 0  .
\end{equation}
In inner coordinates at $O(h^{-1})$ \eqref{equ:two particle plus} reads the same as \eqref{equ:inner first problem} but with no contribution from $P_{\text{out}}^{(0,1)}$ and the addition of the interaction term \eqref{equ:3intin}. The corresponding problem for $P_{\text{in}}^{(1,1)}$ reads
\begin{align}
\label{equ:inner first problem first order phi}
0 =\, & 
	 2\sum_{ e\in B_\v} 
	\Big[
	 P_{\text{in}}^{(1,1)} (\u ,  \v-e,t) - P_{\text{in}}^{(1,1)} (\u ,  \v,t)
	\Big]
	\\
	&+ (\phi_\os p_\os \bfA_{\s,\os} - \phi_\s p_\s \bfA_{\os,\s})(\xin ,t ) \cdot
 	\Big( (2-2 \alpha )
 	\yin \id_{\{ \vert \yin \vert  =1 \}}
 	- \mathpzc{L}^-_\tau  \p_\theta \mathpzc{u} (\{\yin \}) \Big)\nonumber   ,
\end{align}
where we exploit the anti-symmetry of $\mathpzc{u}$ to combine ${\cal E}_{\s, \os, \s}$ and ${\cal E}_{\s, \os, \os}$.
The corresponding boundary condition is obtained by matching to the outer solution,
\begin{equation}
	P_{\text{in}}^{(1,1)} (\u , \v,t) \sim P_{\text{out}}^{(1,1)} (\u , \u , t)  ~\text{ as }~\v \rightarrow \infty   .
\end{equation}
The key observation from Sec. \ref{appendix: dependence} is that the operator $2\sum_{ e\in B_\v}(\tau_{e}+Id)$ acting on a function $f$, is in fact the same as $\mathpzc{L}(0) =(\mathpzc{L}^2_\tau+\mathpzc{L}_0)$ acting on the degree one function of sets $\tw{f}$, where $\tw{f}( \{ \v\} ) = f(\v)$. With this in hand, and using \eqref{equ:u2}, it is easily verifiable  
\begin{equation}
	P_{\text{in}}^{(1,1)}(\u , \v ,t) =  \bfB^{(1,1)}(\u , t)
 	\cdot ( \frac{1}{2}\p^2_\theta \mathpzc{u} (\{\v \}) )
 	-\mathpzc{u}( \{ \v \} ) 
 	+ \bfC^{(1,1)} (\u , t), 
\end{equation}
where 
\begin{align}
	\bfB^{(1,1)} (\u , t) = (\phi_\os p_\os+\phi_\s p_\s) \bfA_{\s,\os}(\xin ,t ),
 && \bfC^{(1,1)} =P_{\text{out}}^{(1,1)} (\u , \u , t) .
 \end{align}

\paragraph*{System of equations for $p_\s$}

Evaluating the two particle interaction term \eqref{equ:inter-explicit}, using the inner solution to $O(h \phi)$, and Taylor expanding about $x$, we find: 
\begin{equation}
	\mathcal{E}_{\s,\os} = \mathcal{E}_{\s,\os}^{(0)} + \phi \mathcal{E}^{(1)}_{\s,\os} + O(h, \phi^2),
\end{equation}
where
\begin{align*}
	\mathcal{E}_{\s,\s'}^{(0)} &= (1+ \alpha ) \nabla \cdot (
	p_\s \nabla p_{\s'} 
	- p_{\s'} \nabla p_\s ), \\
	\mathcal{E}^{(1)}_{\s,\s'} &= (\alpha^{(2)}-\alpha)
	\nabla \cdot \left[
	(\bar \phi_r p_r + \bar \phi_b p_b)(
	p_\s \nabla p_{\s'} 
	- p_{\s'} \nabla p_\s ) \right],
\end{align*}
and $\bar \phi_\s = \phi_\s/\phi$ and $\alpha^{(n)} = \frac{2}{n !} \p_\theta^n \mathpzc{u}( e_k)$.

Rescaling to particle density and taking the limit $h \rightarrow 0$ we find that with this interaction term, equation \eqref{equ:gen int eq} has the gradient flow structure \eqref{equ:gen-hydro}, with free energy \eqref{equ:E0} and mobility \eqref{equ:M-quas} 
where 
\begin{equation*}
	D_s(\phi) = 1 
	- (1+\alpha) \phi
	-(\alpha^{(2)}-\alpha)  \phi^2 +O(\phi^3)  .
\end{equation*}
This agrees with the expansion of $D_s(\phi)$ discussed in Appendix \ref{appendix:Ds}.

\end{appendices}

%\bibliography{asep_biblio.bib}


\end{document}